\documentclass[a4paper,11pt]{article}
\pdfoutput=1 

\usepackage{jheppub_modified} 

\usepackage[T1]{fontenc} 

\usepackage{verbatim}
\usepackage{color}
\usepackage{graphicx}
\usepackage{subcaption}

\usepackage{bigcenter}

\usepackage{sectsty}
\allsectionsfont{\boldmath}

\usepackage{xspace}
\usepackage[dvipsnames,table]{xcolor}
\usepackage{tabularx}
\usepackage[normalem]{ulem}
\newcolumntype{C}[1]{>{\centering\arraybackslash}p{#1}}

\newcommand{\eq}[1]{Eq.~(\ref{#1})}
\newcommand{\bib}[1]{Ref.~\cite{#1}}

\newcommand{\bibs}[1]{\cite{#1}}
\newcommand{\fig}[1]{Fig.~\ref{#1}}
\newcommand{\tab}[1]{Table~\ref{#1}}

\newcommand{\sect}[1]{Section~\ref{#1}}
\newcommand{\ssect}[1]{Subsection~\ref{#1}}

\newcommand{\appen}[1]{Appendix~\ref{#1}}

\newcommand{\bea}{\begin{eqnarray}}
\newcommand{\eea}{\end{eqnarray}}

\newcommand{\crn}{\nonumber \\}

\newcommand{\fr}{\frac}

\newcommand{\gev}{{\unskip\,\text{GeV}}}

\title{Polarized $W^+W^-$ pairs at the LHC: Effects from bottom-quark induced processes at NLO QCD+EW}

\author[a]{Thi Nhung Dao,}
\author[a]{Duc Ninh Le}

\affiliation[a]{Phenikaa Institute for Advanced Study, Phenikaa University, Hanoi 12116, Vietnam}

\emailAdd{nhung.daothi@phenikaa-uni.edu.vn}
\emailAdd{ninh.leduc@phenikaa-uni.edu.vn}

\abstract{We investigate the effects of the bottom-quark induced processes on the doubly polarized cross sections of $W^+W^-$ pair 
production at the LHC. The method to extract the on-shell single-top contribution is provided. Results for phenomenological and 
experimental analyses are given at next-to-leading order (NLO) QCD+EW accuracy, with the leading contribution from the gluon-gluon and photon-photon fusion included. 
We found that the contribution of the bottom-quark induced processes, after the subtraction of the on-shell $tW$ channel, 
is largest for the doubly longitudinal polarization. 
At the integrated cross section level, using a fiducial ATLAS cut with a jet veto, 
the effect is $9\%$ compared to the NLO value of the 
light-quark contribution. 
It increases to $13\%$ after removing the jet veto. 
A bound of the $tW$ interference is calculated for various kinematic distributions, 
showing that this interference effect is, in general, smaller for the no jet veto case. 
Relevant scale uncertainties are calculated to help us decide on the importance of this 
interference.
}

\begin{document}
\maketitle
\flushbottom

\section{Introduction}
\label{sect:intro}
With more and more data, measurements at the 
Large Hadron Collider (LHC) experiments are expanding from unpolarized results to include polarized ones. 
Spin-dependent observables not only help us to test quantum field theory and the Standard Model (SM) 
at a deeper level, but also provide more leverage to search for new physics. 

In this direction, the top quark, the heaviest known particle with spin $1/2$, and the $W^\pm$ and $Z$ bosons, 
the heaviest detected gauge bosons with spin $1$, are the objects of interest. 
Due to their short lifetime (about $10^{-25}$~s for all of them), their intermediate helicity states 
cannot be directly observed. However, different helicity states leave different footprints 
on the kinematic distributions of the decay products, thereby providing us a path toward the separation of different 
helicity contributions.  

Perhaps, more interesting are the spin correlations of these particles when being simultaneously produced. 
Collisions at the LHC provide an ideal, yet realistic, laboratory for these studies. 
Recently, quantum entanglement of the top quark and its anti-partner, an intrinsic quantum property associated with spin, has been observed in 
$t\bar{t}$ production at ATLAS \cite{ATLAS:2023fsd} and CMS \cite{CMS:2024pts}.

Polarization observables in diboson production processes have attracted attention since LEP, where the doubly longitudinal polarization 
fraction was measured for the first time in $W^+ W^-$ pair production \cite{OPAL:1998ixj}. 
The longitudinal polarization is of particular interest because of its connection to the electroweak (EW) symmetry breaking. 

The initial efforts at LEP laid the ground for new studies at the LHC, which fortunately provides us with more options. 
We can now study, not only $W^+ W^-$ pairs but also $ZZ$ and $W^\pm Z$ pairs, for inclusive event selection. 
Even more opportunities open up when $VV^\prime jj$ (with $V,V^\prime=W^\pm, Z$) final states are considered, where same-sign $W^\pm W^\pm$ pairs 
can be produced.  
First measurements of joint-polarized cross sections in $W^\pm Z$, $ZZ$, and 
same-sign $W^\pm W^\pm jj$ have been reported in 
\cite{ATLAS:2022oge,ATLAS:2024qbd} (ATLAS), \cite{ATLAS:2023zrv} (ATLAS), and \cite{CMS:2020etf} (CMS), respectively. 

In parallel, active theoretical efforts to provide more precise theoretical predictions including next-to-leading order (NLO) QCD 
and EW corrections have been going on. Fixed-order results for doubly-polarized cross sections for $ZZ$ \cite{Denner:2021csi}, $W^\pm Z$ \cite{Denner:2020eck,Le:2022lrp,Le:2022ppa}, 
$W^+ W^-$ \cite{Denner:2020bcz,Denner:2023ehn,Dao:2023kwc} have been obtained at NLO including both QCD and EW corrections for fully leptonic decays. 
Next-to-next-to-leading order (NNLO) QCD results for $W^+W^-$ were provided in \cite{Poncelet:2021jmj}, taking into account the light quark induced processes 
(without the bottom quark contribution). 
Semileptonic final state has been considered in \cite{Denner:2022riz} for the case of $WZ$ at NLO QCD. 
NLO QCD+EW corrections have been very recently calculated for the same-sign $W^+W^+jj$ production in \cite{Denner:2024tlu}, marking the first NLO calculation of polarized vector-boson scattering.

Going beyond the fixed order, new results in \cite{Hoppe:2023uux} show that 
it is now possible to simulate polarized events, for multi-boson production processes, 
at the precision level of approximate fixed-order NLO QCD corrections
matched with parton shower using the Monte-Carlo generator SHERPA. 
In addition, the above full NLO QCD calculations in 
$ZZ$ \cite{Denner:2021csi}, $W^\pm Z$ \cite{Denner:2020eck}, 
$W^+ W^-$ \cite{Denner:2020bcz} have been implemented in the 
POWHEG-BOX framework \cite{Pelliccioli:2023zpd}, thereby incorporating parton-shower effects. 
Very recently, polarized $ZZ$ pairs via gluon fusion 
have been generated using the combination of FeynRules and MadGraph5{\_}aMC@NLO \cite{Javurkova:2024bwa}, 
allowing for another option of realistic simulation. 

In this paper, we make additional steps for the $W^+ W^-$ production with fully leptonic decays, 
exploring the interplay with top quark production. 
In the recent works \cite{Denner:2020bcz,Denner:2023ehn,Dao:2023kwc} at NLO, the polarized $W^+ W^-$ pairs from the 
$tW$ are excluded as usually done in ATLAS and CMS analyses, because the $tW$ contribution is large 
and can be separated (to some extent, up to an unkown interference effect usually assumed to be small). 
At NNLO, $W^+ W^-$ pairs can come from the $t\bar{t}$ production channel as well, which overwhelms other 
$W^+ W^-$ production mechanisms. The NNLO QCD calculation in \cite{Poncelet:2021jmj} therefore also excludes 
the $tW$ and $t\bar{t}$ contributions.

In measurements, the top-quark contributions are separated using sophisticated techniques such as 
$b$-tagging, kinematic-separation variables, etc. 
In simulation, additional options are available such as on-shell top quark selection. 
These different techniques complement each other. 
The removal of the $tW$ contribution as done in \cite{Denner:2020bcz,Denner:2023ehn,Dao:2023kwc} (and $t\bar{t}$ in 
\cite{Poncelet:2021jmj}) is too rough, because they removed also $W^+ W^-$ pairs of non-$tW$ origin, 
which are produced by the bottom-quark induced processes. 
In this paper, we use the on-shell technique to remove only the $W^+ W^-$ pairs of the $tW$ origin, where 
the top-quark is required to be on-shell. 
This kind of calculation has been done for the unpolarized events, see e.g. \bib{Tait:1999cf,Frixione:2008yi} for removing the on-shell $t\bar{t}$ 
events in the $tW$ production analysis, and e.g. \bib{Dao:2010nu} for removing the on-shell $tH^-$ events in the 
$W^+H^-$ production analysis in the minimal supersymmetric standard model.  
In this work, the subtraction is done for individual polarized pairs, for the first time (to the best of our knowledge). 

For the case of unpolarized $W^+W^-$ pair production, 
it was pointed out in \cite{Baglio:2013toa} (where one of us is an author, see Section 3.6 there) that the contribution of the bottom-quark induced processes 
is very small at NLO after the subtraction of the on-shell $tW$ production. 
This conclusion is supported by a detailed NNLO QCD calculation \cite{Gehrmann:2014fva}, showing that the 
$b$-induced effect (after the subtraction of the on-shell top processes) is about $2\%$ at $\sqrt{s}=14$~TeV.
This was an additional argument for neglecting the bottom-induced processes.

The context is different this time as we are considering now polarized cross sections, and 
there is a new motivation for studying the bottom-quark induced processes.
It was very recently observed in \cite{Denner:2023ehn,Dao:2023kwc} that the leading-order (LO) $b\bar{b}$ contribution to the 
doubly-longitudinal (LL) polarized cross section is rather large, about $+15\%$ of the NLO QCD prediction (using the 
cut setup of \cite{Dao:2023kwc}). This effect comes from the top-quark mass in the $t$-channel propagator. 
Notice that the $W^+ W^-$ pairs from the LO $b\bar{b}$ annihilation are of non-$tW$ origin. 
Because of this large LO effect, one must consider the NLO corrections. 
At NLO QCD, the $bg$ induced subprocess comes into play, which generates both $tW$ and non-$tW$ originated $W^+ W^-$ pairs 
(the same for $b\gamma$ process in NLO EW corrections). 
Our strategy is first to include the full $bg$ and $b\gamma$ processes as usually done for any NLO calculation. 
In addition, the on-shell $tW$ contribution in these processes are separately calculated. 
After the subtraction of this contribution, we will get the NLO corrections from $b$-quark induced processes 
for the case of non-$tW$ originated $W^+ W^-$ pairs. 

The above approach, which will be adopted in this paper, is well known in the literature as the five-flavor scheme (5FS). 
Alternatively, one can use the four-flavor scheme (4FS) with a massive $b$ quark to calculate the bottom-induced effects. 
In the 4FS, the $b$ quark is treated differently from the other light quarks in the following aspects: 
(i) The bottom parton distribution function (PDF) is zero, hence the $b$ quark is absent in the initial state. 
(ii) A finite value of $m_b$ serves as an infrared (IR) cutoff, allowing for an alternative  
method to explore the IR region. In the 4FS, the $b$-induced effects in $W^+W^-$ pair production 
can be calculated using the $W^+W^-b\bar{b}$ matrix elements. 
This final state includes off-shell $t\bar{t}$, $tW$, and $W^+W^-$ in association with $b\bar{b}$ production 
processes and their interference. As in the case of the 5FS, we have to subtract from this coherent sum 
the on-shell top contribution to obtain the ``top-free'' $W^+W^-$ signal. Unpolarized cross section results for the $W^+W^-b\bar{b}$ production 
in the 4FS can be found in \cite{Frederix:2013gra,Cascioli:2013wga} (at NLO QCD) and 
in \cite{Jezo:2016ujg,Jezo:2023rht} (with parton shower effects). 

To provide experimental colleagues with more information for their analyses, we will present our results for two different analyses: 
with and without the $tW$ contribution. In addition, we will consider two setups: with and without a jet veto. 
It is interesting to notice that ATLAS used a jet veto in their recent analysis \cite{ATLAS:2019rob}, while CMS \cite{CMS:2020mxy} did not. 
An estimate for the bound of the $tW$ interference effects will be provided. 
We will then be able to compare the magnitude of this effect between the two cut setups.

This paper is organized as follows.
In the next section we define the polarized cross sections. 
The new $tW$ contribution is shortly described in \sect{sect:pol_tW}, leaving all 
technical calculation details for \appen{appen_cal_tW}. 
Numerical results are presented in \sect{sect:results}, where integrated cross sections and a few representative 
kinematic distributions are discussed. Additional kinematic distributions important for polarization extraction 
are provided in \appen{appen_dist_add}. 
Finally, we conclude in \sect{sect:conclusion}.
\section{Doubly-polarized cross sections}
\label{sect:pol_xs}
\begin{figure}[ht!]
  \centering
  \includegraphics[width=0.8\textwidth]{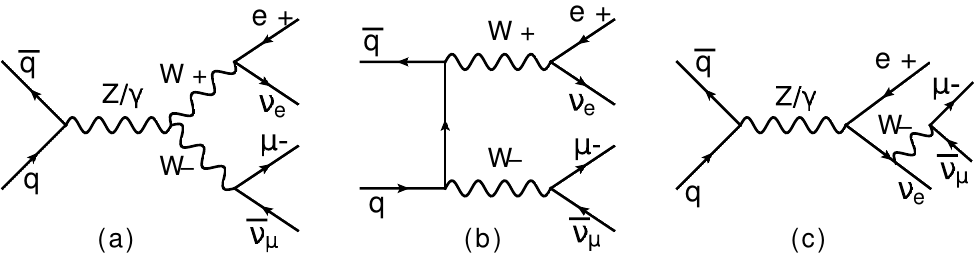}
  \caption{Some representative Feynman diagrams at Born level, including the doubly-resonant contribution (a, b) 
  and non-doubly-resonant ones (c).}
  \label{fig:LO_diags}
\end{figure}

The process of interest reads
\bea
p(k_1) + p(k_2) \to e^{+}(k_3) + \nu_e(k_4) + \mu^{-}
(k_5) + \bar{\nu}_\mu(k_6) + X,
\label{eq:proc1}
\eea
where the final state leptons can be created dominantly from the electroweak gauge bosons, 
namely the $W^\pm$, $Z$, $\gamma$. Some representative Feynman diagrams are shown in \fig{fig:LO_diags}.

The focus of this work is to extract the contribution of the intermediate on-shell (OS) $W^+W^-$ system to this process. 
As in the previous works (see e.g. \cite{Denner:2020bcz,Denner:2020eck,Denner:2021csi,Le:2022ppa,Denner:2023ehn,Dao:2023kwc}), we use the double-pole approximation (DPA) \cite{Aeppli:1993cb,Aeppli:1993rs,Denner:2000bj} 
to define this contribution. 

The idea of the DPA is to select only the doubly-resonant diagrams (i.e. diagrams (a) and (b) in \fig{fig:LO_diags}). 
However, these diagrams do not form a gauge invariant group in the general case of off-shell momenta. 
The second ingredient of the DPA is then an on-shell mapping, which defines a set of on-shell momenta from the off-shell ones. 
The DPA amplitudes, a product of the $W^+W^-$ production amplitude and the 
$W^+ \to e^+ \nu_e$, $W^- \to \mu^- \bar{\nu}_\mu$ decay amplitudes, are calculated using the on-shell momenta. 
The results are gauge invariant because the individual production and decay processes are gauge invariant. 

More specifically, the DPA amplitudes are defined as, at leading order (LO):
\bea
\mathcal{A}_\text{LO,DPA}^{\bar{q}q\to V_1V_2\to 4l} = \fr{1}{Q_1Q_2}
\sum_{\lambda_1,\lambda_2=1}^{3}
\mathcal{A}_\text{LO}^{\bar{q}q\to V_1V_2}(\hat{k}_i,\lambda_1,\lambda_2)\mathcal{A}_\text{LO}^{V_1\to
    l_1l_2}(\hat{k}_i,\lambda_1)\mathcal{A}_\text{LO}^{V_2\to l_3l_4}(\hat{k}_i,\lambda_2)
,\label{eq:LO_DPA}
\eea
with 
\bea
Q_j = q_j^2 - M_{V_j}^2 + iM_{V_j}\Gamma_{V_j}\;\; (j=1,2),
\label{eq:Qi_def}
\eea
where $q_1 = k_3+k_4$, $q_2 = k_5 + k_6$, $M_{V_j}$ and $\Gamma_{V_j}$ are the
physical mass and width of the gauge boson $V_j$, and $\lambda_j$ are the
polarization indices of the gauge bosons. The helicity indices of the initial-state quarks and final-state leptons are implicit, 
meaning that the full helicity amplitude on the l.h.s. $\mathcal{A}_\text{LO,DPA}^{\bar{q}q\to V_1V_2\to 4l}$ 
stands for 
$\mathcal{A}_\text{LO,DPA}^{\bar{q}q\to V_1V_2\to 4l}(\sigma_{\bar{q}},\sigma_q,\sigma_{l_1},\sigma_{l_2},\sigma_{l_3},\sigma_{l_4})$ 
with $\sigma_{\bar{q}}$, $\sigma_q$ 
being the helicity indices of the initial-state quarks; $\sigma_{l_1}$, $\sigma_{l_2}$, $\sigma_{l_3}$, $\sigma_{l_4}$
of the final-state leptons. 
Correspondingly, on the r.h.s. we have $\mathcal{A}_\text{LO}^{\bar{q}q\to V_1V_2} = \mathcal{A}_\text{LO}^{\bar{q}q\to V_1V_2}(\sigma_{\bar{q}},\sigma_q)$, $\mathcal{A}_\text{LO}^{V_1\to
    l_1l_2}=\mathcal{A}_\text{LO}^{V_1\to
    l_1l_2}(\sigma_{l_1},\sigma_{l_2})$, 
    $\mathcal{A}_\text{LO}^{V_2\to l_3l_4}=\mathcal{A}_\text{LO}^{V_2\to l_3l_4}(\sigma_{l_3},\sigma_{l_4})$. 
    The squared amplitude then automatically includes correlations between different helicity states of the final leptons. 
    This is the key difference between the DPA and the narrow-width approximation where spin correlations are neglected. 

The momenta denoted with a hat, $\hat{k}_i$, are the OS momenta, obtained via an OS mapping as above mentioned.   
This mapping is not unique, however the differences between different choices are very small, 
of order $\alpha \Gamma_V/(\pi M_V)$ \cite{Denner:2000bj}, hence of no practical importance. 
In this work, we use the same OS mappings as in \bib{Denner:2021csi,Dao:2023kwc}. 
A necessary condition for the existence of OS mappings is that the 
invariant mass of the four-lepton system must be greater than $2M_W$, 
which is required at LO and NLO. 

As usual, from \eq{eq:LO_DPA}, we then separate the unpolarized cross section into the LL, LT, TL, TT, and interference 
terms. Here, L and T mean longitudinal and transverse polarization modes, respectively. 
The transverse polarization is the coherent sum (i.e. interference is included) 
of the transverse-left and transverse-right polarizations.  

For the NLO QCD and EW corrections, the definition of double-pole amplitudes 
need include the virtual corrections, the gluon/photon induced and radiation processes. 
This issue has been fully discussed in \bib{Denner:2023ehn,Dao:2023kwc}, hence there is no need to repeat it here.

The new contribution of this work is to calculate the NLO QCD and EW corrections to the subprocess $\bar{b}b \to W^+W^- \to e^+\nu_e \mu^-\bar{\nu}_\mu$, where, differently from the similar processes of the first two generations (which have been calculated in \bib{Denner:2023ehn,Dao:2023kwc}), there is a $tW$ contribution occurring at NLO. This is the topic of the next section.

\section{The $tW$ contribution}
\label{sect:pol_tW}
\begin{figure}[ht!]
  \centering
  \includegraphics[width=0.8\textwidth]{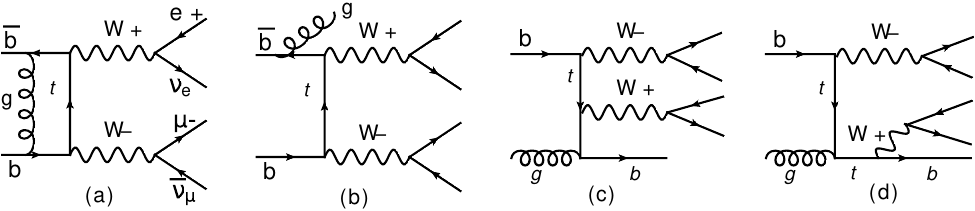}
  \caption{Bottom-quark induced processes at NLO QCD in the 5FS: virtual corrections (a), real-gluon emission (b) and bottom emission (c,d), where (d) is the $tW$ contribution. Similar diagrams occur for the NLO EW corrections, where the gluon is replaced by the photon.}
  \label{fig:bb_NLO_diags}
\end{figure}

It was observed in \bib{Denner:2023ehn,Dao:2023kwc} that the LO $b\bar{b}$ contribution is rather large for the LL cross section, 
about $15\%$ of the NLO cross section \cite{Dao:2023kwc}. This poses a question on the size of the NLO corrections of this subprocess. 

The NLO QCD corrections are divided into two groups: $b\bar{b}$ and $bg$ induced corrections. 
The former includes the virtual and real gluon emission contributions (groups (a) and (b) in \fig{fig:bb_NLO_diags}), while the latter has an extra $b$-quark emission in the final state (groups (c) and (d) in \fig{fig:bb_NLO_diags}). Similar classification is done for the NLO EW corrections, where the gluon is replaced by the photon (or $W$/$Z$ for the virtual corrections).  

In this work, we use the 5FS, where the bottom quark is approximated as massless (except for 
the loop-induced gluon-gluon fusion where we take $m_b = 4.7$~GeV) and 
the bottom PDF of the proton is non-vanishing. 
At NLO, the bottom quark induced processes include all processes with at least one $b$ quark in the initial state, as shown 
in \fig{fig:bb_NLO_diags}. 
At NNLO, where two additional jets are possible, processes with two $b$ quarks in the final state 
such as $gg \to W^+W^- b\bar{b}$ and $q\bar{q} \to W^+W^- b\bar{b}$ (with $q=u,d,c,s$) are also counted as 
$b$-induced processes. 

By now, in most ATLAS and CMS analyses of the $W^+W^-$ process, the top contribution is part of the background. 
Subtracting the $t$ backgrounds will remove both the resolved and unresolved bottom quark contribution of the 
top origin. At NLO, the signal includes the $WW$ diagrams ((a), (b), and (c) of \fig{fig:bb_NLO_diags}) and 
the interference between diagrams (c) and (d) of \fig{fig:bb_NLO_diags}.

This interference term cannot be calculated in a gauge-invariant way \footnote{A method was proposed in \bib{Frixione:2008yi} to 
estimate the interference effect from the difference between the on-shell subtraction and diagram removal results. This is 
not gauge invariant because the diagram removal result is not gauge invariant.}. However, its bound can be estimated as follows.
Since this interference effect occurs only in the $bg$ (for QCD corrections) and $b\gamma$ (EW corrections) processes, 
we consider only the $bg$ and $b\gamma$ processes ($b$ here can be the bottom or anti-bottom quark). 
We first calculate the full (including $WW$, $tW$ and their interference) result,
 then subtract from this the on-shell $tW$ contribution. 
There remains the sum of the $WW$ and the interference contributions, named $\hat{\sigma}_\text{TW-int}$ in the following (the hat symbol is to indicate this is a bound). 
Since there is no method to separate the interference term from the $WW$ one 
in a gauge-invariant way, we therefore
interpret the absolute value of $\hat{\sigma}_\text{TW-int}$ as the bound of the interference effect. 
We will keep the sign to show that it can be negative. 
The bound $\hat{\sigma}_\text{TW-int}$ is gauge invariant by construction.

More specifically, for the NLO QCD contribution we have
\bea
\hat{\sigma}_\text{TW-int} = \sigma_{bg} - \sigma_{bg}^{tW},
\eea
where the $b$ and $g$ PDFs are included and the second term is the on-shell $tW$ contribution detailed in \appen{appen_cal_tW}. 
$\sigma_{bg}$ is the cross section of the process $bg \to e^+ \nu_e \mu^- \bar{\nu}_\mu b$. 

The $bg \to e^+ \nu_e \mu^- \bar{\nu}_\mu b$ amplitude includes the diagrams (c) and (d) of \fig{fig:bb_NLO_diags} and 
all related diagrams which form a gauge invariant group. The diagrams like \fig{fig:bb_NLO_diags} (c) 
are IR divergent because of the singular $g\to b\bar{b}$ splitting. 
As usual, this divergence is absorbed into the PDF as explained in \cite{Catani:1996vz} for the case of hadron-hadron collision. 

The cross section $\sigma_{bg}$ 
is calculated using the dipole-subtraction method \cite{Catani:1996vz,Dittmaier:1999mb}. 
It includes a contribution 
from the so-called $(n+1)$-particle part (where the extra $b$-radiation amplitudes and subtraction terms are combined) and the corresponding 
PK-operator term (where the integrated part and PDF counterterms are included). 
These two contributions are individually finite and hence can be separately calculated using an IR regularization, for which 
we choose the mass regularization \cite{Dittmaier:1999mb}. 
Similarly, the $tW$ interference bound for the NLO EW contribution can be 
straightforwardly obtained.

In this work, we have to calculate the on-shell $tW$ contribution to the individual 
polarized cross sections. To achieve this, one has to implement a proper on-shell mapping, 
to ensure that the intermediate top quark and the two $W$ bosons 
are all on-shell. Details of this new piece of calculation are provided in \appen{appen_cal_tW}. 

It is worth noting that \bib{Denner:2023ehn} provided results for the full off-shell unpolarized NLO EW cross sections 
including the $tW$ contribution. Since the DPA results are very close to the full off-shell ones, a rough comparison will 
be possible for the case of unpolarized cross section. 
\section{Numerical results}
\label{sect:results}
We employ here the identical input parameters and renormalization schemes (for QCD and EW corrections) 
as in our previous investigations \cite{Le:2022ppa, Dao:2023pkl}. Note that $m_b=4.7\gev$ for the loop-induced gluon gluon contribution, 
otherwise the bottom quark is massless. The top-quark parameters are $m_t = 173\gev$, $\Gamma_t = 1.42\gev$. 
We discuss numerical results for the LHC at $13$ TeV center-of-mass energy. 
The factorization and renormalization scales are chosen at a fixed value 
$\mu_F = \mu_R = M_W$, where $M_W=80.385\,\gev$. 
For the parton distribution functions, we use the Hessian set 
{\tt
  LUXqed17\char`_plus\char`_PDF4LHC15\char`_nnlo\char`_30}~\bibs{Watt:2012tq,Gao:2013bia,Harland-Lang:2014zoa,Ball:2014uwa,Butterworth:2015oua,Dulat:2015mca,deFlorian:2015ujt,Carrazza:2015aoa,Manohar:2016nzj,Manohar:2017eqh} via the library {\tt LHAPDF6}~\bibs{Buckley:2014ana}. For the PDF counterterms, the DIS scheme (see e.g. \cite{Dittmaier:2009cr}) 
  has been used in the calculation of the NLO EW corrections. 
  We have checked that the difference compared to the $\overline{\text{MS}}$ scheme (see e.g. \cite{Dittmaier:2009cr}) is 
  negligible. For the NLO QCD corrections, the $\overline{\text{MS}}$ scheme is used.   
  
For subprocesses with a real photon emission,  we 
do lepton-photon recombination to define a dressed lepton before applying real 
analysis cuts. Momentum of a dressed lepton 
is defined as $p'_\ell = p_\ell + p_\gamma$ if $\Delta
R(\ell,\gamma) \equiv \sqrt{(\Delta\eta)^2+(\Delta\phi)^2}< 0.1$, i.e. when the photon 
is close enough to the bare lepton.   
In case two charged leptons meet this requirement, the one nearest to the photon is selected. The only 
photons that can undergo this recombination are those with $|y_\gamma|<5$; otherwise, they are 
considered as to have been lost in the beam pipe. 
The letter $\ell$ can be either $e$ or $\mu$ 
and $p$ here denotes a momentum vector in the laboratory (LAB) frame.

As in \bibs{Denner:2020bcz,Dao:2023kwc}, we apply  the ATLAS fiducial phase-space cuts which are defined as follows. \\
\underline{YesVeto setup:}
\begin{align}
        & p_{T,\ell} > 27\gev, \quad p_{T,\text{miss}} > 20\gev, \quad |\eta_\ell|<2.5, \quad m_{e\mu} > 55\gev,\crn
        & \text{jet veto (no jets with $p_{T,j}>35\gev$ and $|\eta_j|<4.5$)}.
\end{align}
Note that the jet here, in the context of a NLO calculation, can be a light quark, a bottom quark or a gluon. 
This set of cuts  has been adapted from the ATLAS analysis \cite{ATLAS:2019rob}.\\
\underline{NoVeto setup:}\\ 
The same cut setup as above, but without the jet veto.\\

A jet veto is usually used to reduce the top-quark backgrounds, with the cost of an increase in the theoretical prediction 
uncertainty \cite{Stewart:2011cf}. In the CMS $W^+W^-$ analyses presented in \bib{CMS:2020mxy}, a jet veto is not applied.  
In addition to these cuts, the requirement of $m_{4l} > 2M_W$, as above mentioned, 
is imposed in our code to ensure that the events with two on-shell $W$ bosons are available. 
Moreover, for the on-shell $tW$ contribution, another cut of $m_{b4l} > m_t + M_W$ (invariant mass of the four leptons and the 
bottom quark) is required, see \eq{eq:OS_cut_tW_tW}.

Numerical results for polarized cross sections will be presented for the $WW$ center-of-mass frame, called $WW$ frame for 
short. Here, we define a few quantities before presenting our finding. Since 
 our goal is to examine
 the corrections resulting from the bottom-quark driven processes, we therefore 
isolate the bottom contributions from the other ones as: 
\begin{itemize} 
 \item  $ \sigma_\text{NoB} $ is defined as the total of the cross sections 
 from the light-quark ($q=u,d,c,s$) induced processes as well as $gg$ and $\gamma\gamma$ processes. The light-quark induced cross sections incorporate  both the NLO QCD and NLO EW corrections calculated in \cite{Dao:2023kwc}. The $gg$ and $\gamma\gamma$ contributions are computed at LO only (see to \cite{Dao:2023kwc}). 
We note that the NLO corrections to the $\gamma\gamma$ are very small for all the polarized cross sections, 
hence can be safely neglected. 
\item $\sigma_\text{YesTW}$ includes  $ \sigma_\text{NoB} $ and the total cross section at NLO QCD and EW
from the bottom-quark induced processes. 
\item $\sigma_\text{NoTW}$ is equal to $\sigma_\text{YesTW}$ after subtracting the on-shell $tW$ contribution. 
\item $\hat{\sigma}_\text{TW-int}$, defined in \ssect{sect:pol_tW}, is the bound of the $tW$ interference. 
\end{itemize}
In the following subsections, numerical results for the integrated polarized cross sections and their differential ones are discussed. 

Before that, a few words on the comparison with \cite{Denner:2023ehn} are of benefit for the reader. 
Without the bottom-quark induced processes, the integrated cross sections 
of this work (which are identical to \cite{Dao:2023kwc}) agree very well with ones of 
\cite{Denner:2023ehn} at LO, with the differences all less than $0.07\%$ for all polarized cases, using the numerical 
input of \cite{Denner:2023ehn}.
Including the NLO EW corrections, our NLO results are a bit smaller, with the differences (with respect to \cite{Denner:2023ehn}) 
of $-0.01\%$, $-0.1\%$, $-0.1\%$, $-0.4\%$, and $-0.3\%$ 
for the LL, LT, TL, TT, and unpolarized cases, respectively. 
We do not yet understand the origin of the $-0.4\%$ difference for the TT case. 
However, this tiny discrepancy is completely negligible compared to other sources of uncertainties, e.g. the scale uncertainties. 
Taking into account the bottom-quark induced contribution, \bib{Denner:2023ehn} provides NLO EW results only for the case of full off-shell 
unpolarized cross section. They obtained $+2.54\%$ for the NLO EW correction (including the $tW$ contribution), 
while our corresponding DPA result reads $+2.61\%$. This level of agreement is satisfactory, given that the difference between the DPA and 
the full off-shell results is about $-3.5\%$ at LO and $-3.3\%$ at NLO EW, using the setup of \bib{Denner:2023ehn} (see Table 1 there).

The numerical results of this work are obtained using our in-house computer program MulBos (MultiBoson production), which 
has been used for our previous papers \cite{Le:2022lrp,Le:2022ppa,Dao:2023kwc}. 
The ingredients of this program include the helicity amplitudes for the production and decay processes, generated by 
FeynArt \cite{Hahn:2000kx} and FormCalc \cite{Hahn:1998yk}, an in-house library for one-loop integrals named LoopInts. 
The tensor one-loop integrals are calculated using the standard technique of Passarino-Veltman reduction \cite{Passarino:1978jh}, 
while the scalar integrals are computed as in \cite{'tHooft:1978xw, Nhung:2009pm, Denner:2010tr}.  
The phase space integration is done using the Monte-Carlo integrator BASES \cite{Kawabata:1995th}, with the help of 
useful resonance mapping routines publicly available in VBFNLO \cite{Baglio:2024gyp}. 
For other details of the NLO QCD+EW calculations, the reader is referred to \cite{Le:2022ppa,Dao:2023kwc}.      

\subsection{Integrated polarized cross sections}
\label{sect:XS}
We first present results for the unpolarized and four polarized (LL, LT, TL, TT) integrated cross sections in \tab{tab:xs_fr_all_YesVeto}. 
The polarization interference (denoted Pol-int), calculated by subtracting the polarized cross sections from the unpolarized one, is shown 
in the bottom row. Besides the three defined cross sections,  $ \sigma_\text{NoB} $, $\sigma_\text{YesTW}$
 and $\sigma_\text{NoTW}$, with their statistical and scale uncertainties, we show some ratios. 
We choose to normalize new effects to the $\sigma_\text{NoB}$ to quantify various bottom-induced corrections. 
The LO $b\bar{b}$ contributions is quantified by a R-factor, 
 $R^\text{LO}_{b\bar{b}} = (\sigma_\text{NoB}+ \sigma_{b\bar{b}}^{\text{LO}})/\sigma_\text{NoB}$
 for unpolarized and each polarized cross sections. Similarly for the NLO contribution to the 
 b-induced processes, we  give also the R-factors defined as:
 $R_\text{NoTW} = \sigma_\text{NoTW}/\sigma_\text{NoB}$; $R_\text{YesTW} = \sigma_\text{YesTW}/\sigma_\text{NoB}$. 
 The interference effect is quantified by $\hat{\delta}_\text{TW-int}=\hat{\sigma}_\text{TW-int}/\sigma_\text{NoB}$. 
  The last three columns of \tab{tab:xs_fr_all_YesVeto}, we give 
 the corresponding polarization fractions $f_{X,i}=\sigma_{X,i}/\sigma_{X,\text{Unpol.}}$ with $X=$ $\text{NoB}$, $\text{NoTW}$, $\text{YesTW}$ and $i=$ Unpol., LL, LT, TL, TT, Pol-int.

In \tab{tab:xs_fr_all_YesVeto}, the scale uncertainties are calculated using the usual seven-point method. The 
central scale is $\mu_F = \mu_R = \mu_0 = M_W$. The seven scale points are $(\mu_F, \mu_R) = (i\mu_0, j\mu_0)$ 
with $i=0.5,1,2$ and $j=0.5,1,2$, with the constraint of $i/j < 4$ and $j/i < 4$ to avoid wide-separated scales. 
The relative uncertainties are then defined as $\Delta_{-} = [\min_{i,j}(\sigma(i\mu_0, j\mu_0)) - \sigma(\mu_0, \mu_0)]/\sigma(\mu_0, \mu_0)$ 
and $\Delta_{+} = [\max_{i,j}(\sigma(i\mu_0, j\mu_0)) - \sigma(\mu_0, \mu_0)]/\sigma(\mu_0, \mu_0)$. 
These scale uncertainties are provided for the NLO cross sections. 
\begin{table}[h!]
 \renewcommand{\arraystretch}{1.3}
\begin{bigcenter}
\setlength\tabcolsep{0.03cm}
{\fontsize{6.0}{6.0}
\begin{tabular}{|c|c|c|c|c|c|c|c|c|c|c|}\hline
  & $\sigma_\text{NoB}\,\text{[fb]}$ & $\sigma_\text{NoTW}\,\text{[fb]}$ & $\sigma_\text{YesTW}\,\text{[fb]}$ & $R^\text{LO}_{b\bar{b}}$ & $R_\text{NoTW}$ & $R_\text{YesTW}$ & $\hat{\delta}_\text{TW-int}\,\text{[\%]}$ & $f_\text{NoB}\,\text{[\%]}$ & $f_\text{NoTW}\,\text{[\%]}$ & $f_\text{YesTW}\,\text{[\%]}$ \\
\hline
{\fontsize{6.0}{6.0}$\text{Unpol.}$} & $218.47(3)^{+2.2\%}_{-2.1\%}$ & $220.50(3)^{+2.1\%}_{-2.0\%}$ & $266.12(3)^{+3.7\%}_{-3.8\%}$ & $1.02$ & $1.01$ & $1.22$ & $-0.75$ & $100$ & $100$ & $100$\\
\hline
{\fontsize{6.0}{6.0}$W^{+}_{L}W^{-}_{L}$} & $14.34^{+1.8\%}_{-2.6\%}$ & $15.59^{+1.2\%}_{-2.2\%}$ & $29.88^{+6.3\%}_{-6.7\%}$ & $1.15$ & $1.09$ & $2.08$ & $-4.41$ & $6.6$ & $7.1$ & $11.2$\\
{\fontsize{6.0}{6.0}$W^{+}_{L}W^{-}_{T}$} & $24.79^{+1.9\%}_{-2.5\%}$ & $25.31^{+1.6\%}_{-2.5\%}$ & $34.74^{+4.4\%}_{-5.2\%}$ & $1.04$ & $1.02$ & $1.40$ & $-1.64$ & $11.3$ & $11.5$ & $13.1$\\
{\fontsize{6.0}{6.0}$W^{+}_{T}W^{-}_{L}$} & $25.47^{+2.1\%}_{-2.5\%}$ & $25.99^{+1.8\%}_{-2.4\%}$ & $35.42^{+4.5\%}_{-5.1\%}$ & $1.04$ & $1.02$ & $1.39$ & $-1.59$ & $11.7$ & $11.8$ & $13.3$\\
{\fontsize{6.0}{6.0}$W^{+}_{T}W^{-}_{T}$} & $152.59(3)^{+2.2\%}_{-1.9\%}$ & $152.67(3)^{+2.2\%}_{-1.9\%}$ & $166.19(3)^{+3.0\%}_{-2.7\%}$ & $1.00$ & $1.00$ & $1.09$ & $-0.19$ & $69.8$ & $69.2$ & $62.5$\\
\hline
{\fontsize{6.0}{6.0}$\text{Pol-int}$} & $1.27(4)$ & $0.93(4)$ & $-0.12(4)$ & $--$ & $--$ & $--$ & $--$ & $0.6$ & $0.4$ & $-0.0$\\
\hline
\end{tabular}
}
\caption{\small Integrated unpolarized and doubly polarized cross sections in fb
  calculated in the $WW$ frame for the process $p p \to W^+ W^-\to e^+ \nu_e \mu^- \bar{\nu}_\mu + X$ with the YesVeto setup.   
  The statistical uncertainties (in parenthesis) are displayed on the final
  digits of the central prediction when significant. Seven-point scale
  uncertainty is also provided for the cross sections as sub- and
  superscripts in percent. }
\label{tab:xs_fr_all_YesVeto}
\end{bigcenter}
\end{table}

The results are interesting. We found that the bottom-quark induced correction at NLOQCDEW to the unpolarized cross section is 
small, being $+1\%$, when the single-top contribution has been subtracted, confirming the result of \bib{Baglio:2013toa}. 
This correction is also very small 
for the TT cross section, which is the largest contribution to the unpolarized cross section. 
The correction is largest for the LL cross section, being $+9\%$, which 
is significantly different from the LO prediction of $+15\%$ found in \cite{Dao:2023kwc,Denner:2023ehn}. 
The LO $b\bar{b}$ results are also shown in \tab{tab:xs_fr_all_YesVeto} so that we can appreciate the changes from 
LO to NLOQCDEW. For the TL and LT cases, the NLO corrections are the same, of about $+2\%$. 

If the single-top contribution is included in the analysis, then the values of the unpolarized and polarized cross 
sections change drastically. The bottom-quark induced correction now reads $+108\%$, $+40\%$, $+39\%$, $+9\%$ 
for the LL, LT, TL, TT cross sections, respectively, leading to a large correction of $+22\%$ for the unpolarized case. 

These large corrections naturally lead us to the question on the size of the $tW$ interference effects. 
As discussed in \sect{sect:pol_tW}, this interference term cannot be calculated, but its bound (named $\hat{\delta}_\text{TW-int}$) 
can be computed and is shown in \tab{tab:xs_fr_all_YesVeto}. We found that $\hat{\delta}_\text{TW-int} = -4.42\%$ for the LL cross section, 
and significantly smaller (in absolute value) for the other polarizations. This effect is rather large and would have an impact 
on the measurement of the LL polarized cross section, if the single-top contribution is subtracted.   
\begin{table}[h!]
 \renewcommand{\arraystretch}{1.3}
\begin{bigcenter}
\setlength\tabcolsep{0.03cm}
\fontsize{8.0}{8.0}
\begin{tabular}{|c|c|c|c|c|c|c|c|}\hline
  & $\sigma^\text{LO}_{b}\,\text{[fb]}$ & $\sigma^\text{NoTW}_{b}\,\text{[fb]}$ & $\sigma^\text{YesTW}_{b}\,\text{[fb]}$ & $\sigma^\text{NoTW}_{bg}\,\text{[fb]}$ & $\sigma^\text{YesTW}_{bg}\,\text{[fb]}$ & $\sigma^\text{NoTW}_{b\gamma}\,\text{[fb]}$ & $\sigma^\text{YesTW}_{b\gamma}\,\text{[fb]}$ \\
\hline
{\fontsize{6.0}{6.0}$\text{Unpol.}$} & $3.94$ & $2.03(1)$ & $47.65(1)$ & $-1.62(1)$ & $42.66(1)$ & $-0.01$ & $1.34$\\
\hline
{\fontsize{6.0}{6.0}$W^{+}_{L}W^{-}_{L}$} & $2.12$ & $1.25$ & $15.54$ & $-0.63$ & $13.50$ & $-0.00$ & $0.16$\\
{\fontsize{6.0}{6.0}$W^{+}_{L}W^{-}_{T}$} & $0.96$ & $0.52$ & $9.95$ & $-0.40$ & $8.84$ & $-0.00$ & $0.17$\\
{\fontsize{6.0}{6.0}$W^{+}_{T}W^{-}_{L}$} & $0.96$ & $0.52$ & $9.95$ & $-0.40$ & $8.85$ & $-0.00$ & $0.17$\\
{\fontsize{6.0}{6.0}$W^{+}_{T}W^{-}_{T}$} & $0.36$ & $0.07$ & $13.60$ & $-0.29$ & $12.45$ & $-0.00$ & $0.78$\\
\hline
{\fontsize{6.0}{6.0}$\text{Interf.}$} & $-0.46$ & $-0.34(1)$ & $-1.39(1)$ & $0.11(1)$ & $-0.98(1)$ & $0.00$ & $0.04$\\
\hline
\end{tabular}
\caption{\small Integrated cross sections at LO and NLO for the $b$ induced processes for the YesVeto setup. 
The QCD $bg$ and EW $b\gamma$ induced processes are separately provided for a better understanding of the origin of the $tW$-interference effects. Very small values are rounded to zero at the chosen precision level.}
\label{tab:xs_fr_b_YesVeto}
\end{bigcenter}
\end{table}

In \tab{tab:xs_fr_b_YesVeto} we show in more detail the bottom-quark induced contributions at LO and NLO. 
The NLO results are calculated in two scenarios: with and without single-top contribution. 
The bound of the single-top interference $\hat{\sigma}_\text{TW-int} = \sigma^\text{NoTW}_{bg}+\sigma^\text{NoTW}_{b\gamma}$ is 
now split into QCD and EW parts. The results show that the QCD $bg$ process is completely dominant, while the EW 
contribution $\sigma^\text{NoTW}_{b\gamma}$ is negligible. The surprising result here is that the LL cross section is 
largest (in absolute value) for the $bg$ process. This explains why the single-top interference is so large for the LL polarization.

\begin{table}[h!]
 \renewcommand{\arraystretch}{1.3}
\begin{bigcenter}
\setlength\tabcolsep{0.03cm}
{\fontsize{6.0}{6.0}
\begin{tabular}{|c|c|c|c|c|c|c|c|c|c|c|}\hline
  & $\sigma_\text{NoB}\,\text{[fb]}$ & $\sigma_\text{NoTW}\,\text{[fb]}$ & $\sigma_\text{YesTW}\,\text{[fb]}$ & $R^\text{LO}_{b\bar{b}}$ & $R_\text{NoTW}$ & $R_\text{YesTW}$ & $\hat{\delta}_\text{TW-int}\,\text{[\%]}$ & $f_\text{NoB}\,\text{[\%]}$ & $f_\text{NoTW}\,\text{[\%]}$ & $f_\text{YesTW}\,\text{[\%]}$ \\
\hline
{\fontsize{6.0}{6.0}$\text{Unpol.}$} & $327.94(4)^{+5.4\%}_{-4.2\%}$ & $334.17(4)^{+5.4\%}_{-4.1\%}$ & $620.13(4)^{+8.3\%}_{-6.5\%}$ & $1.01$ & $1.02$ & $1.89$ & $0.62$ & $100$ & $100$ & $100$\\
\hline
{\fontsize{6.0}{6.0}$W^{+}_{L}W^{-}_{L}$} & $18.68^{+4.1\%}_{-3.3\%}$ & $21.04(1)^{+4.0\%}_{-2.9\%}$ & $83.66(1)^{+9.9\%}_{-9.5\%}$ & $1.11$ & $1.13$ & $4.48$ & $1.04$ & $5.7$ & $6.3$ & $13.5$\\
{\fontsize{6.0}{6.0}$W^{+}_{L}W^{-}_{T}$} & $43.33^{+6.0\%}_{-4.9\%}$ & $44.86(1)^{+6.1\%}_{-4.8\%}$ & $110.18(1)^{+9.5\%}_{-8.1\%}$ & $1.02$ & $1.04$ & $2.54$ & $1.12$ & $13.2$ & $13.4$ & $17.8$\\
{\fontsize{6.0}{6.0}$W^{+}_{T}W^{-}_{L}$} & $44.22(1)^{+6.2\%}_{-4.9\%}$ & $45.77(1)^{+6.2\%}_{-4.8\%}$ & $111.06(1)^{+9.5\%}_{-8.1\%}$ & $1.02$ & $1.03$ & $2.51$ & $1.12$ & $13.5$ & $13.7$ & $17.9$\\
{\fontsize{6.0}{6.0}$W^{+}_{T}W^{-}_{T}$} & $221.43(3)^{+5.3\%}_{-4.1\%}$ & $222.80(3)^{+5.3\%}_{-4.1\%}$ & $321.82(3)^{+7.2\%}_{-5.6\%}$ & $1.00$ & $1.01$ & $1.45$ & $0.43$ & $67.5$ & $66.7$ & $51.9$\\
\hline
{\fontsize{6.0}{6.0}$\text{Pol-int}$} & $0.28(5)$ & $-0.30(5)$ & $-6.60(5)$ & $--$ & $--$ & $--$ & $--$ & $0.1$ & $-0.1$ & $-1.1$\\
\hline
\end{tabular}
}
\caption{\small The same as \tab{tab:xs_fr_all_YesVeto}, but for the NoVeto setup.}
\label{tab:xs_fr_all_NoVeto}
\end{bigcenter}
\end{table}
\begin{table}[h!]
 \renewcommand{\arraystretch}{1.3}
\begin{bigcenter}
\setlength\tabcolsep{0.03cm}
\fontsize{8.0}{8.0}
\begin{tabular}{|c|c|c|c|c|c|c|c|}\hline
  & $\sigma^\text{LO}_{b}\,\text{[fb]}$ & $\sigma^\text{NoTW}_{b}\,\text{[fb]}$ & $\sigma^\text{YesTW}_{b}\,\text{[fb]}$ & $\sigma^\text{NoTW}_{bg}\,\text{[fb]}$ & $\sigma^\text{YesTW}_{bg}\,\text{[fb]}$ & $\sigma^\text{NoTW}_{b\gamma}\,\text{[fb]}$ & $\sigma^\text{YesTW}_{b\gamma}\,\text{[fb]}$ \\
\hline
{\fontsize{6.0}{6.0}$\text{Unpol.}$} & $3.93$ & $6.23(2)$ & $292.19(2)$ & $1.91(2)$ & $278.89(2)$ & $0.13$ & $9.11$\\
\hline
{\fontsize{6.0}{6.0}$W^{+}_{L}W^{-}_{L}$} & $2.12$ & $2.36(1)$ & $64.98(1)$ & $0.18$ & $62.07(1)$ & $0.01$ & $0.75$\\
{\fontsize{6.0}{6.0}$W^{+}_{L}W^{-}_{T}$} & $0.96$ & $1.53(1)$ & $66.85(1)$ & $0.46(1)$ & $64.62(1)$ & $0.02$ & $1.18$\\
{\fontsize{6.0}{6.0}$W^{+}_{T}W^{-}_{L}$} & $0.96$ & $1.54(1)$ & $66.83(1)$ & $0.48(1)$ & $64.60(1)$ & $0.02$ & $1.18$\\
{\fontsize{6.0}{6.0}$W^{+}_{T}W^{-}_{T}$} & $0.36$ & $1.38(1)$ & $100.40(1)$ & $0.88(1)$ & $94.30(1)$ & $0.08$ & $5.68$\\
\hline
{\fontsize{6.0}{6.0}$\text{Interf.}$} & $-0.46$ & $-0.58(3)$ & $-6.88(3)$ & $-0.09(2)$ & $-6.70(3)$ & $-0.00$ & $0.31$\\
\hline
\end{tabular}
\caption{\small The same as \tab{tab:xs_fr_b_YesVeto}, but for the NoVeto setup.}
\label{tab:xs_fr_b_NoVeto}
\end{bigcenter}
\end{table}

The same results for the NoVeto setup are shown in \tab{tab:xs_fr_all_NoVeto} and \tab{tab:xs_fr_b_NoVeto}. 
Compared to the YesVeto setup, the NLO unpolarized cross sections $\sigma_\text{NoB}$, $\sigma_\text{NoTW}$, $\sigma_\text{YesTW}$ 
increase by a factor of $1.5$, $1.5$, $2.3$, respectively. 
These factors differ for different polarizations, being 
$1.3$, $1.4$, $2.8$ for LL; $1.7$, $1.8$, $3.2$ for LT; 
$1.7$, $1.8$, $3.1$ for TL; 
$1.5$, $1.5$, $1.9$ for TT. It is interesting to notice that the $tW$ interference for the LL case changes from $-4.4\%$ to $+1.1\%$ when the jet veto is removed. 

In conclusion, from the integrated cross section results, we see that, in order to keep all the interference effects small, 
the best choice is the NoVeto setup. This conclusion will be supported by almost all differential distributions, except for the 
case of the $\cos(\theta_e^{WW})$, as will be shown in the next section.
Including the $tW$ contribution changes the polarization fractions (the last three columns of \tab{tab:xs_fr_all_YesVeto} and 
\tab{tab:xs_fr_all_NoVeto}) violantly, in particular for the LL case.

\subsection{Kinematic distributions}
\label{sect:dist}
\begin{figure}[t!]
  \centering
  \begin{tabular}{cc}
  \includegraphics[width=0.48\textwidth]{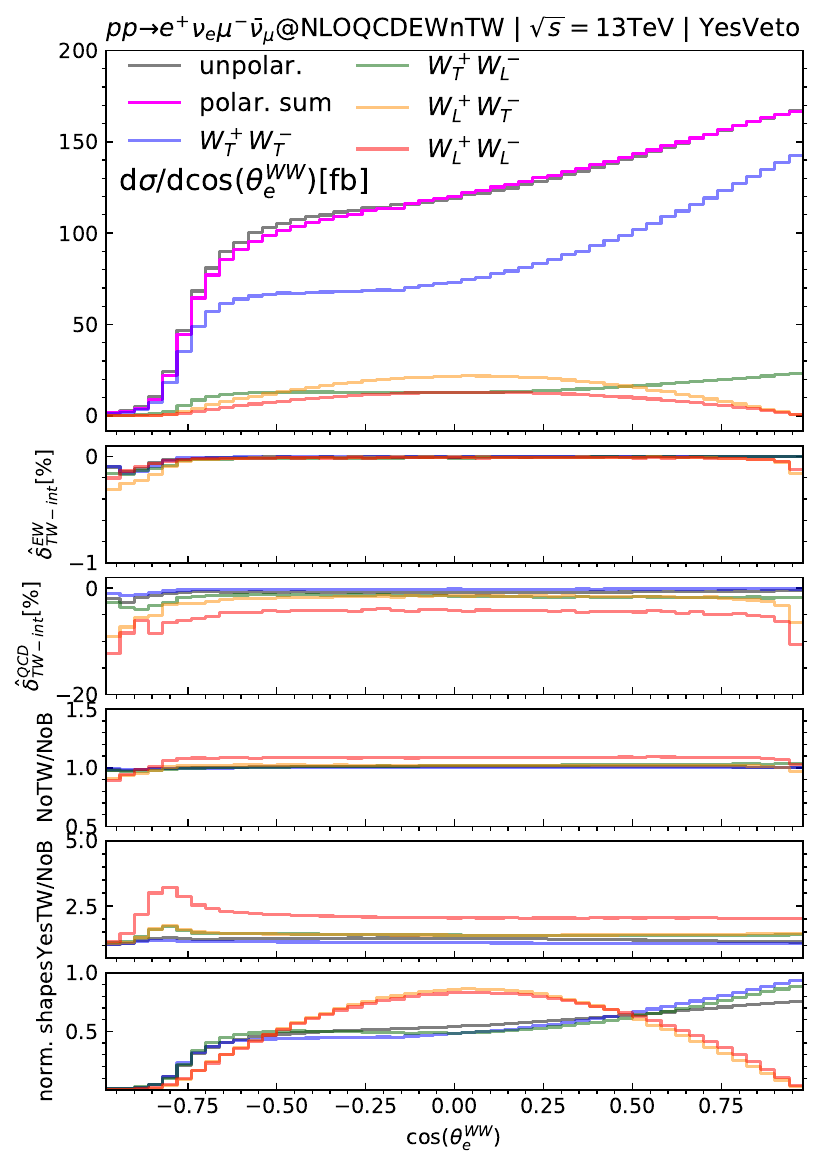} 
  \includegraphics[width=0.48\textwidth]{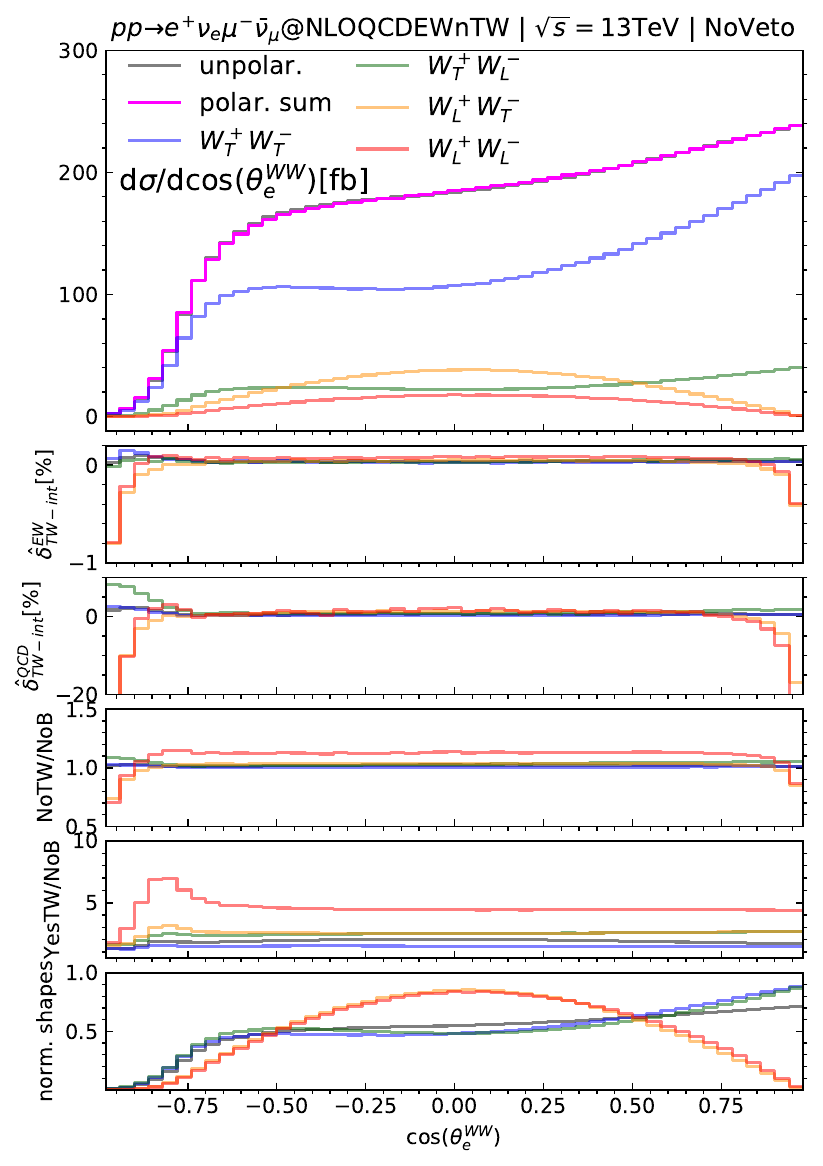}\\
  \includegraphics[width=0.48\textwidth]{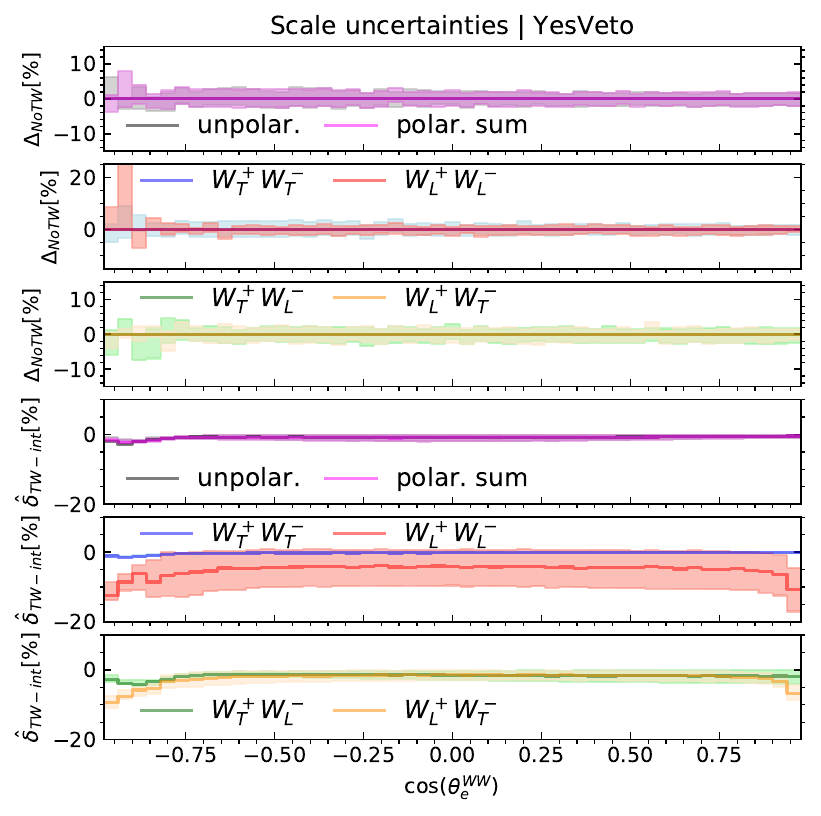} 
  \includegraphics[width=0.48\textwidth]{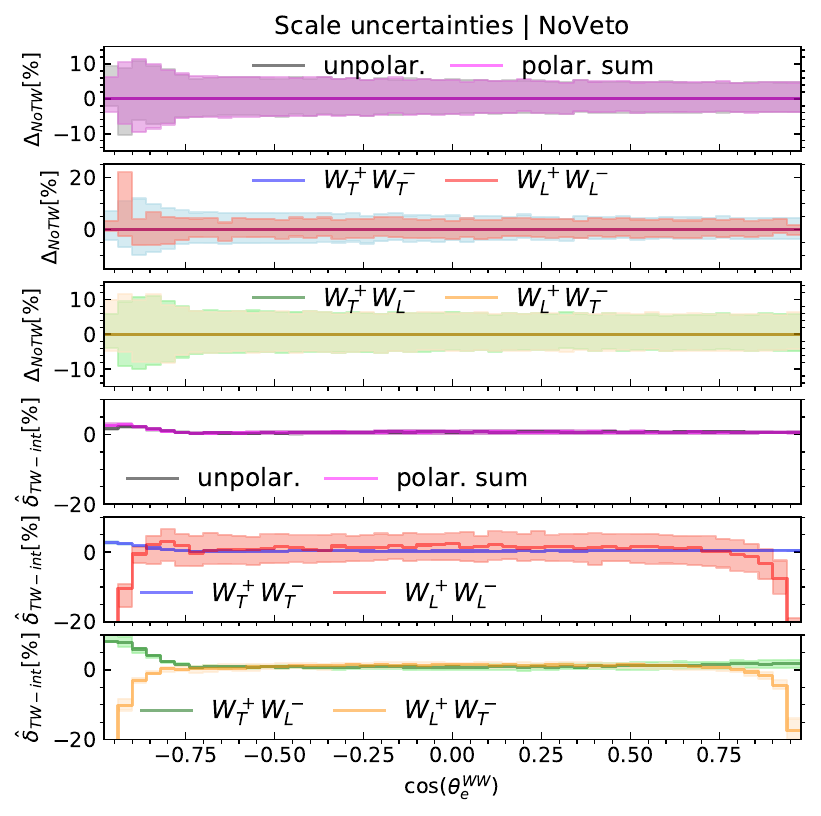}
  \end{tabular}
  \caption{Distributions in $\cos\theta^{WW}_{e^{+}}$ for the YesVeto (left) and
    NoVeto (right) setups. Details are provided in the main text.}
  \label{fig:dist_costheta_e_VV}
\end{figure}
\begin{figure}[t!]
  \centering
  \begin{tabular}{cc}
  \includegraphics[width=0.48\textwidth]{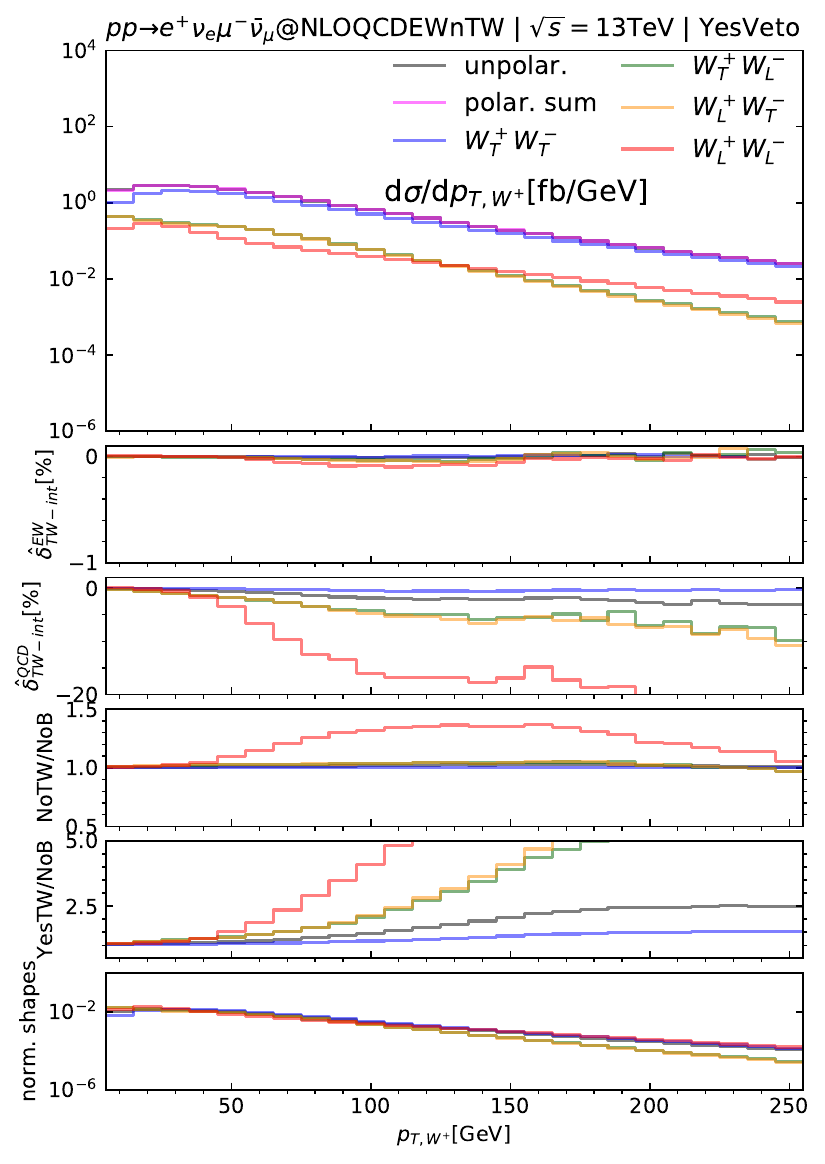} 
  \includegraphics[width=0.48\textwidth]{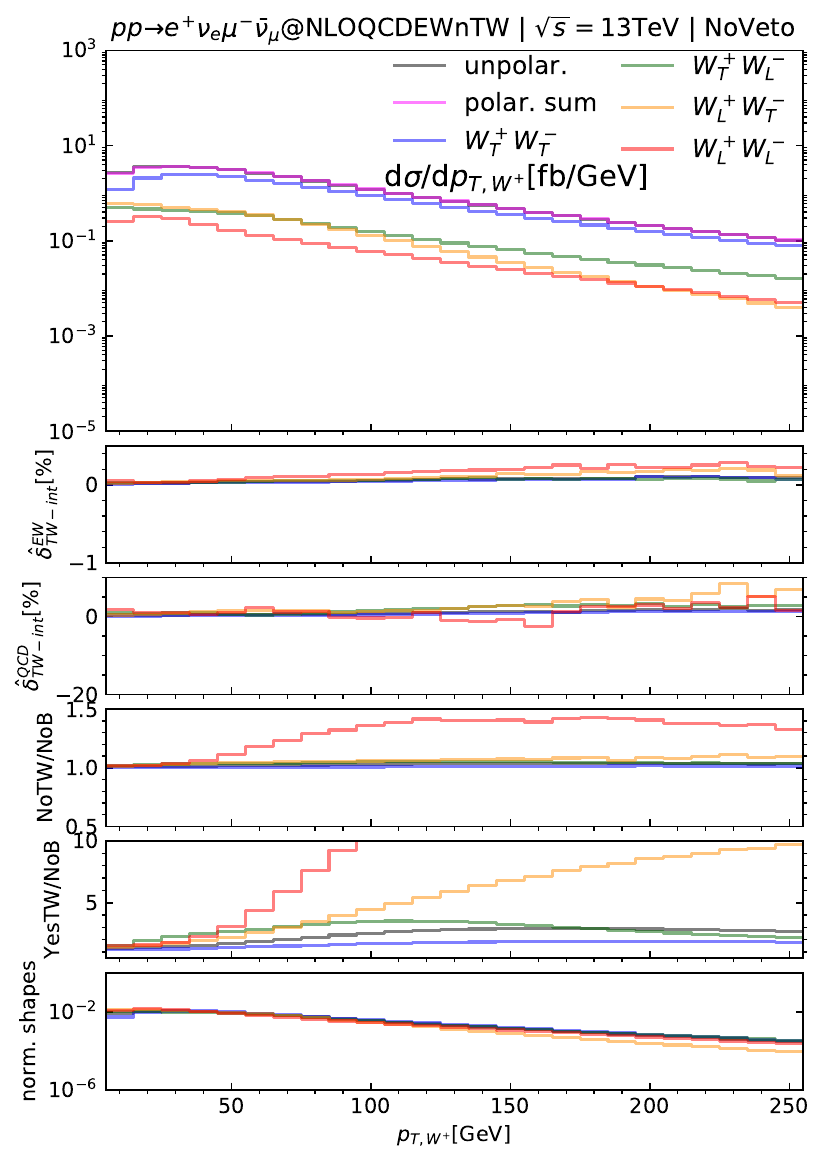}\\
  \includegraphics[width=0.48\textwidth]{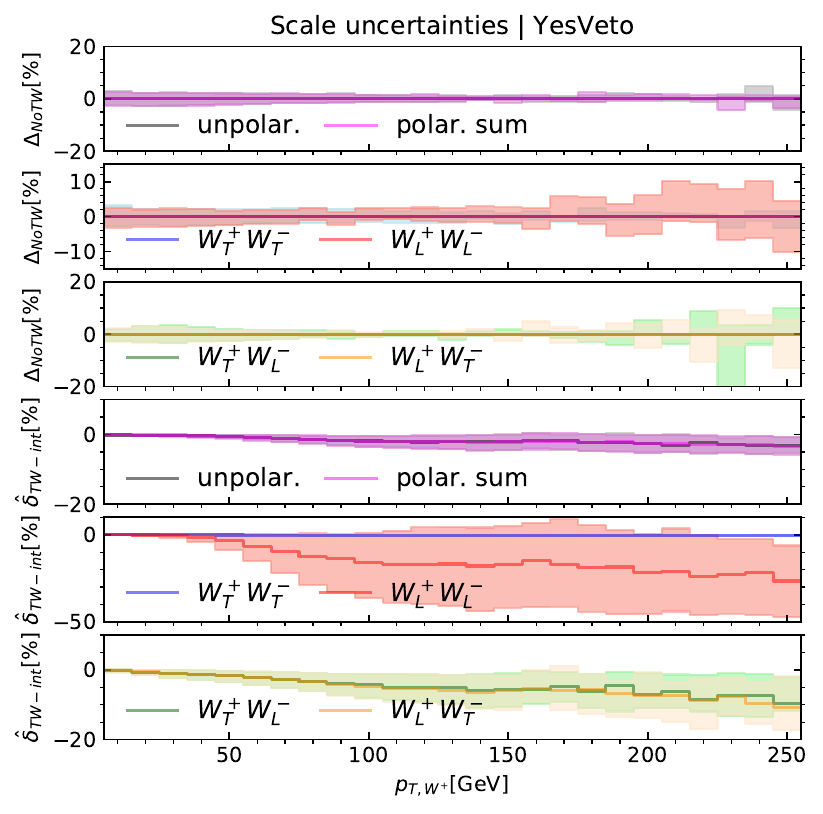} 
  \includegraphics[width=0.48\textwidth]{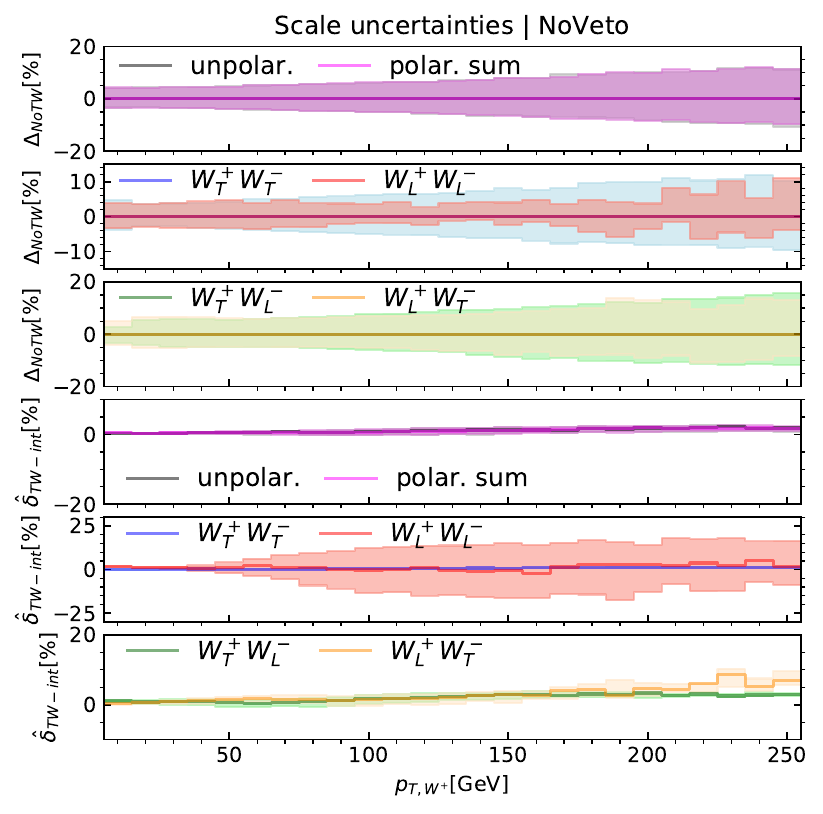} 
  \end{tabular}
  \caption{Same as \fig{fig:dist_costheta_e_VV} but for the transverse momentum of the $W^+$.}
  \label{fig:dist_pT_W1}
\end{figure}
The effects from $b$-quark induced processes on the doubly longitudinal polarization are also interesting when
considering differential cross sections. 
We first discuss two important distributions, an angular distribution 
of the electron and a transverse momentum distribution of the $W^+$, 
in \fig{fig:dist_costheta_e_VV} and \fig{fig:dist_pT_W1}, respectively.  
Other distributions Figs.~\ref{fig:dist_cos35}-\ref{fig:dist_pT_e}, which are also important for polarization separation, 
are shown in \appen{appen_dist_add}.

For each distribution we display four plots arranged in two columns and two rows. 
The first column is for the YesVeto setup, the second one NoVeto. 
Each plot on the first row has six panels. 
The big panel shows the values of the NoTW unpolarized
and polarized cross sections, which include all contributions except the 
on-shell $tW$ channel. 
The small panels, from top to bottom, are the EW $tW$-interference bound, 
the QCD $tW$-interference bound, the ratio of $\sigma_\text{NoTW}/\sigma_\text{NoB}$, 
the ratio of $\sigma_\text{YesTW}/\sigma_\text{NoB}$,
the normalized shapes of the distributions shown in the big panel.
Each plot on the second row has six panels showing the scale uncertainties. 
From top to bottom, the first three panels show the relative scale uncertainties 
of the NoTW cross sections provided in the big panel on the first row. 
The next three panels display the scale uncertainty band of the EW+QCD $tW$ interference bound 
$\hat{\delta}_\text{TW-int}=\hat{\sigma}_\text{TW-int}/\sigma_\text{NoB}$. 
This band is calculated by varying the scales $\mu_F$ and $\mu_R$ in $\hat{\sigma}_\text{TW-int}$ 
while $\sigma_\text{NoB}$ is fixed at the central scale $\mu_F = \mu_R = M_W$. 
Same as for the integrated cross sections, the seven-scale method is used to compute the scale uncertainty band.

In \fig{fig:dist_costheta_e_VV}, we present the distributions in $
\cos(\theta_e^\text{WW})$. 
Here $\theta_e^\text{WW}$ is defined as the angle between the momentum vectors 
of the positron ($\vec{p}_{e^+}$) and of the $W^+$ boson ($\vec{p}_{W^+}$). 
The positron momentum is defined in the 
rest frame of the $W^+$ boson, while the $W^+$ momentum in the $WW$ frame. 

From top to bottom, in the big panel the TT, TL, LT, LL polarized differential cross sections 
together with their sum (labeled polar. sum) are plotted. 
The unpolarized result is also shown in grey. 
The difference between the unpolarized and the 
polarization sum (pink) lines gives the interference between the TT, TL, LT and LL amplitudes. 
This polarization interference is small for both YesVeto and NoVeto setups, 
 and is smaller for the latter. 
 It is however not always small, see e.g. the $\Delta\phi_{e,\mu}$ distribution in \fig{fig:dist_Dphi}.

In the next two small panels, we show the $tW$-interference effect ($\hat{\sigma}_{\text{TW-int}}$) 
of the EW and QCD contributions separately. 
As expected, the QCD contribution is completely dominant over the whole range. 
For the YesVeto case, both EW and QCD interference bounds
are mostly negative across the whole range of $\cos(\theta_e^\text{WW})$. 
The bound is largest for the LL polarization, ranging from $-15\%$ to $-5\%$, being most pronounced at 
the edges. 
The distributions are very different for the NoVeto case. 
Both EW and QCD bounds 
  can be positive and move closer to zero for $|\cos(\theta_e^\text{WW})| < 0.75$ (see the QCD LL in particular). 
  However, at the two edges, the magnitude of the interference bounds becomes somewhat larger 
  for the LL and LT cases.
 For the TT and TL polarizations, the QCD bounds are positive at the two edges.
For the LL case, we see clearly a cancellation between the negative and positive contributions 
in different regions, explaining the small and positive value of 
the $tW$-interference bound for the integrated cross section of the NoVeto setup. 
 
 The two following panels display the two respective R-factors ($R_\text{NoTW} $, $R_\text{YesTW}$)
 as function of $\cos(\theta_e^\text{WW})$, showing the bottom-induced effects in comparison 
 to the contribution of the two first generations. As seen from the NoTW panels for the YesVeto setup, 
 the R-factor of the
 TT polarization is nearly flat and close to $1$ across the range. 
 It is more interesting for
 the other three polarizations, especially the LL one. 
 The $R_\text{NoTW} $ is less than $1$ for $\cos(\theta_e^\text{WW})<-0.85$ 
 since the $tW$ interference bound is more negative in this region.
 It is however slightly over $1$ for the rest. 
 The picture is completely different for the YesTW plot. 
 The LL R-factor is mostly greater than $2$ 
 and stands out from the other polarizations. 
 Similarly for the case of NoVeto setup, the R-factors of the doubly longitudinal polarization
 are even larger than the corresponding ones of the YesVeto case. 
 Its maximum reaches $7$ around $\cos(\theta_e^\text{WW})=-0.8$ for the YesTW case.
 
In the next panels of the $\cos(\theta_e^\text{WW})$ distribution, we
 display the normalized
distributions from the top panel where the integrated cross
sections are all normalized to unity, to highlight the shape
differences. We see that the LT and LL polarizations are completely 
different from the TL and TT ones. This gives additional power to distinguish polarized signals. 

The relative scale uncertainties of the NoTW differential cross sections plotted in the top panel 
are displayed in the next three panels. 
The remarkable feature is the difference between the YesVeto and NoVeto results, 
showing that the scale uncertainties of the former are significantly smaller for all polarizations. 
We note that the scale uncertainties in both cases are calculated in the same way, being straightforwardly obtained 
from the results of the seven scale points. 
This result is also reflected in the integrated cross sections shown in \tab{tab:xs_fr_all_YesVeto} and \tab{tab:xs_fr_all_NoVeto}. 
Scale uncertainties for the YesVeto setup are also provided in \cite{Denner:2020bcz} for the integrated cross 
sections using a different PDF set (the NNPDF3.1). 
For the NoVeto case, the reader can consult the $WZ$ results in \cite{Denner:2020eck,Le:2022lrp}, 
which are expected to be close to the $W^+W^-$ ones.
These results are quite close to the values in \tab{tab:xs_fr_all_YesVeto} and \tab{tab:xs_fr_all_NoVeto}. 

The smallness of the scale uncertainties in the YesVeto case is due to a cancellation between 
the inclusive zero-jet cross section and the inclusive one-jet cross section, as shown in 
\cite{Stewart:2011cf}. The magnitude of this cancellation depends on the value of $p_{T,\text{veto}}$, which 
is 35 GeV in this work. It turns out that the cancellation is very strong around this value 
(see Fig. 1 (bottom left) of \cite{Stewart:2011cf}). 

An alternative method to calculate the scale uncertainties for the YesVeto 
case is using the quadrature sum $\sqrt{(\Delta\sigma_{N_j\ge 0})^2+(\Delta\sigma_{N_j\ge 1})^2}$ 
where $N_j$ is the number of jets, as proposed in \cite{Stewart:2011cf}. 
The underlying assumption here is that the two scale uncertainties 
$\Delta\sigma_{N_j\ge 0}$ and $\Delta\sigma_{N_j\ge 1}$ are independent. 
The YesVeto scale uncertainties calculated in this way are always larger than the 
NoVeto ones. Care must therefore be taken when using the YesVeto 
scale uncertainties presented in this work for comparison with data, as the values 
may be underestimated. This is a well-known issue of the jet-veto method. 

The last three panels show the scale uncertainties of the QCD+EW $tW$-interference bound. 
The interesting feature here is that the scale uncertainty bands cover the line 
of vanishing interference for all distributions (including the ones in \appen{appen_dist_add}). 
The uncertainty-band width increases with the magnitude of the interference bound. 
Noticeable exception is the edges of the $\cos(\theta_e^\text{WW})$ distribution for the LL case. 
However, the scale uncertainties of the signal are large in these regions (see the $\Delta_\text{NoTW}$ panel). 
This result shows that the $tW$-interference is, at the current level of precision, 
consistent with zero within the scale uncertainties. Note that, the scale uncertainties of the 
$tW$-interference bound are calculated here at LO. 
The importance of this interference effect will become clearer when the scale uncertainties are significantly reduced 
by including higher-order corrections.

We now discuss the transverse momentum distributions of the $W^+$ in \fig{fig:dist_pT_W1}. The distributions 
are presented similarly to the $\cos(\theta_e^\text{WW})$ distributions, with YesVeto setup on the left and NoVeto 
setup on the right. The $tW$-inteference effects  of the EW contributions  ($\hat {\sigma}_{\text{TW-int}}^{\text{EW}}$)
 for all polarizations are not interesting, they are close to zero and show a slight dependence on $p_{T,W^+}$. However
 $\hat {\sigma}_{\text{TW-int}}^{\text{QCD}}$ in the YesVeto  case is more  interesting. While the TT polarization does not
 depend on  $p_{T,W^+}$, the mixed polarizations (TL, LT) show a similar dependence and the LL polarization exhibits a strongest
 dependence. In particular, the QCD contribution to the $tW$-inteference bound for the LL polarization is negative and, 
 starting from  $p_{T, W^+}> 50\,\gev$, falls down rapidly before reaching $-20\%$ at $p_{T, W^+} \approx 200\,\gev$. 
 This behavior of 
 $\hat {\sigma}_{\text{TW-int}}^{\text{QCD}}$ does not occur for the NoVeto setup. The $tW$-interference bound is 
 much smaller there. The smallness of this interference in the NoVeto case is also visible for all 
 distributions displayed in \appen{appen_dist_add}. This result suggests that the NoVeto setup seems to be a better choice 
 for precise polarization measurements.  
 
 The dependence of the two R-factors ($R_\text{NoTW} $, $R_\text{YesTW}$) for the LL polarization is also interesting. The
 $R_\text{NoTW}$ line curves down and peaks around $p_{T, W^+}= 125$ GeV ($150$ GeV) for YesVeto (NoVeto), while the 
 $R_\text{YesTW}$ (including the on-shell $tW$ contribution) increases fast with $p_{T, W^+}$ for both cut setups.

Finally, we have a technical note here. To have a smooth plot for the $tW$-interference is not easy as it is the result of a subtraction 
 of two huge numbers whose values are very close. One can see this from the fluctuations in the distributions, in particular \fig{fig:dist_pT_W1} and \fig{fig:dist_pT_e}. In order to obtain the shapes as shown in this paper, we have doubled the 
 statistics (for both individual runs and the number of random seeds) for the components contributing to the $tW$-interference. 
 Specifically, $20$ random seeds, each with $2^{24}$ random points (for the $(n+1)$ QCD components, $2^{22}$ for the corresponding EW) 
 have been used to calculate the $tW$-interference distributions. Usually, choosing $10$ random seeds with $2^{23}$ QCD points is enough to have nice distributions.      
\section{Conclusions}
\label{sect:conclusion}
In this work, we have calculated the bottom-quark induced contribution to the doubly polarized cross sections 
of $W^+W^-$ pair production at the LHC using a fully leptonic decay mode at NLO QCD+EW level in the five-flavor scheme. 

After the subtraction of the on-shell $tW$ production, this contribution is very small ($1\%$ for the YesVeto setup, $2\%$ for NoVeto)
for the integrated unpolarized cross section, in agreement with the result of \bibs{Baglio:2013toa,Gehrmann:2014fva}. 
However, this effect is not equal on the four polarized cross sections. 
While being negligible for the TT polarization, it is largest for the LL case, at $9\%$ ($13
\%$) for the YesVeto (NoVeto) setup. 
Beyond the integrated results, 
differential cross sections for individual polarizations have been presented and discussed,
showing a distinct shape of the bottom-quark induced contribution to the LL polarization in some distributions. 
The magnitude of this effect can be large, e.g. reaching $30\%$ at $p_{T,W} \approx 100$~GeV. 
These results show that the LL polarization is more sensitive to the third-generation quark contribution. 

For completeness, the corresponding results keeping the $tW$ contribution have also been provided. 
As expected, the polarization fractions change drastically compared to the $tW$-subtracted analysis, in particular for the 
LL mode. As above, the kinematic distributions show a distinct shape and more pronounced magnitude of the bottom and top quark effects 
on the LL polarization.

In addition, we have computed a bound of the $tW$ interference and found that it can reach $-4.4\%$ ($+1.0\%$) for the integrated LL cross section 
for the YesVeto (NoVeto) setup. 
For the kinematic distributions presented in this paper, this interference effect is negligible for NLO EW corrections, 
being less than $1\%$ in absolute value. 
For NLO QCD case, it can however be very large at some phase space regions, reaching $-20\%$ level for the LL polarization.  
Remarkably, we found that the $tW$ interference effect is significantly smaller for the NoVeto setup in comparison to the YesVeto one, for almost all differential distributions, except for the case of $\cos(\theta_e^{WW})$ angular distribution where the interference is large at the edges for both cut setups. 

Scale uncertainties on the signal and on the $tW$-interference bound have been provided using the direct scale variation method. 
For the $tW$-interference bound, we found that the width of the scale-uncertainty band increases with the magnitude of the bound. 
Except for the edges of the $\cos(\theta_e^{WW})$ distribution, the uncertainty band covers the case of vanishing $tW$-interference. 
In addition, at the $|\cos(\theta_e^{WW})|=1$ edges, the scale uncertainties on the signal are large, overwhelming 
the interference effect. It therefore remains hard to decide on the importance of the $tW$ interference for polarization measurements.
 

\appendix

\section{Details of the $tW$ calculation}
\label{appen_cal_tW} 
As discussed in \sect{sect:pol_tW}, the $tW$ contribution occurs at NLO QCD in the process $bg \to e^+ \nu_e \mu^- \bar{\nu}_\mu b$ 
and at NLO EW in the process $b\gamma \to e^+ \nu_e \mu^- \bar{\nu}_\mu b$. Here the notation $b$ stands for either the bottom or anti-bottom quark. 
These contributions can be calculated in a gauge invariant way by requiring that three particles, including a top quark and two $W$ bosons, in the intermediate  states  must be on-shell. 
The on-shell $tW$ process then reads (we show the $bg$ process as an example, the $b\gamma$ calculation is similar)
{\fontsize{9.0}{9.0}
\bea
b (k_1) + g (k_2) \to t (p_t) W^{-} (p_{W^-}) \to W^{+} (p_{W^+}) W^{-} (p_{W^-}) b \to 
e^+ (k_3) \nu_e (k_4) \mu^- (k_5) \bar{\nu}_\mu (k_6) b (k_7).
\eea
}

The unpolarized amplitude at LO then reads (The DPA is applied for the $tW^-$ production. An additional $W^+$ pole occurs due to the subsequent top-quark decay.):
\begin{align}
\mathcal{A}_\text{LO,DPA}^{bg\to tW^{-} \to 4lb} = & \fr{1}{Q_t Q_{W^+} Q_{W^-}}\sum_{\lambda_1,\lambda_2=1}^{3}{\big(}\sum_{s_t=1}^{2}
[\mathcal{A}_\text{LO}^{bg\to tW^{-}}(\hat{k}_i,s_t,\lambda_2)\mathcal{A}_\text{LO}^{t\to W^{+} b}(\hat{k}_i,s_t,\lambda_1)]\crn
&
[\mathcal{A}_\text{LO}^{W^{+}\to e^{+} \nu_e}(\hat{k}_i,\lambda_1)
\mathcal{A}_\text{LO}^{W^{-}\to \mu^{-} \bar{\nu}_\mu}(\hat{k}_i,\lambda_2)
]{\big)},
\label{eq:A_dpa_tw}
\end{align}
where $s_t$, $\lambda_1$, $\lambda_2$ are the helicity indices of the top quark, $W^+$, $W^-$, respectively. As usual, the 
propagator factor is calculated using the off-shell momenta as
\bea
Q_t = p_t^2 - m_t^2 + im_t\Gamma_t, \quad Q_j = p_j^2 - M_{W}^2 + iM_{W}\Gamma_{W}\,\, (j=W^+,W^-), 
\eea
with $p_{W^+} = k_3 + k_4$, $p_{W^-} = k_5 + k_6$, $p_t = p_{W^+} + k_7$. 

In \eq{eq:A_dpa_tw}, it is important to notice that all the helicity amplitude factors must be calculated using the on-shell 
momenta denoted with a hat, $\hat{k}_i$ (with $i=1,7$). These momenta satisfy the following constraints:
\bea
\hat{k}_i^2 = 0, \;
(\hat{k}_3 + \hat{k}_4)^2 = (\hat{k}_5 + \hat{k}_6)^2 = M_W^2, \;
(\hat{k}_3 + \hat{k}_4 + \hat{k}_7)^2 = m_t^2,
\eea
which is solvable when the off-shell momenta pass these cuts
\begin{align}
(k_3 + k_4 + k_5 + k_6)^2 &> 4M_W^2,\label{eq:OS_cut_tW_WW}\\
(k_3 + k_4 + k_5 + k_6 + k_7)^2 &> (m_t + M_W)^2.
\label{eq:OS_cut_tW_tW}
\end{align}

The on-shell momenta $\hat{k}_i$ are then computed as follows. 
In the first step, the on-shell momenta for the $bg\to tW^-$ process are calculated using 
Eqs. (A.1) and (A.2) of Ref.~\cite{Baglio:2018rcu} (which followed \cite{Denner:2000bj}). The key point here is that the $\hat{p}_t$ 
and $\hat{p}_{W^-}$ are calculated in the $tW$ center-of-mass system, and the spatial direction of 
$\hat{p}_t$ is chosen to be the same as the original off-shell direction, i.e. $\vec{\hat{p}}_t = c\vec{p}_t$ 
with $c$ being a real number.

In the second step, the on-shell momenta of the two decays $t (\hat{p}_t) \to e^+ (\bar{k}_3) \nu_e (\bar{k}_4) b (\bar{k}_7)$ 
and $W^- (\hat{p}_{W^-}) \to \mu^- (\hat{k}_5) \bar{\nu}_\mu (\hat{k}_6)$ are calculated 
using the mapping described in \cite{Denner:2021csi}. 
The decay-product momenta satisfy
\begin{align}
(\bar{k}_3 + \bar{k}_4 + \bar{k}_7)^2 &= m_t^2,\\
(\hat{k}_5 + \hat{k}_6)^2 &= M_W^2.
\end{align}
At the final step, the these momenta must be given in the the $tW$ center-of-mass system. 

Notice that the momenta of the top-decay product are denoted with a bar, not a hat. This is because the lepton momenta do not satisfy the 
on-shell $W$ constraint, namely $(\bar{k}_3 + \bar{k}_4)^2 \neq M_W^2$. 
To impose this additional constraint, we then boost the momenta 
$\bar{k}_3$, $\bar{k}_4$, $\bar{k}_7$ to the top-quark rest frame, then split the final 
state into two particles:
\bea
\bar{p}_{W^+} = \bar{k}_3 + \bar{k}_4,\quad 
\bar{p}_b = \bar{k}_7.
\eea
We then repeat the above first step to obtain the on-shell momenta 
$\hat{p}_{W^+}$ and $\hat{p}_b$. 
The on-shell momenta of the decay $W^+ (\hat{p}_{W^+}) \to e^+ (\hat{k}_{3}) \nu_e (\hat{k}_{4})$ are then 
calculated as described in the second step. Finally, these momenta are boosted from the top-quark rest frame to the $tW$ center-of-mass system. 

From \eq{eq:A_dpa_tw}, we then select the $W^+_L W^-_L$, $W^+_L W^-_T$, $W^+_T W^-_L$, $W^+_T W^-_T$ squared amplitudes to calculate 
the polarized cross sections.  

\section{Additional kinematic distributions}
\label{appen_dist_add} 
\begin{figure}[t!]
  \centering
  \begin{tabular}{cc}
  \includegraphics[width=0.48\textwidth]{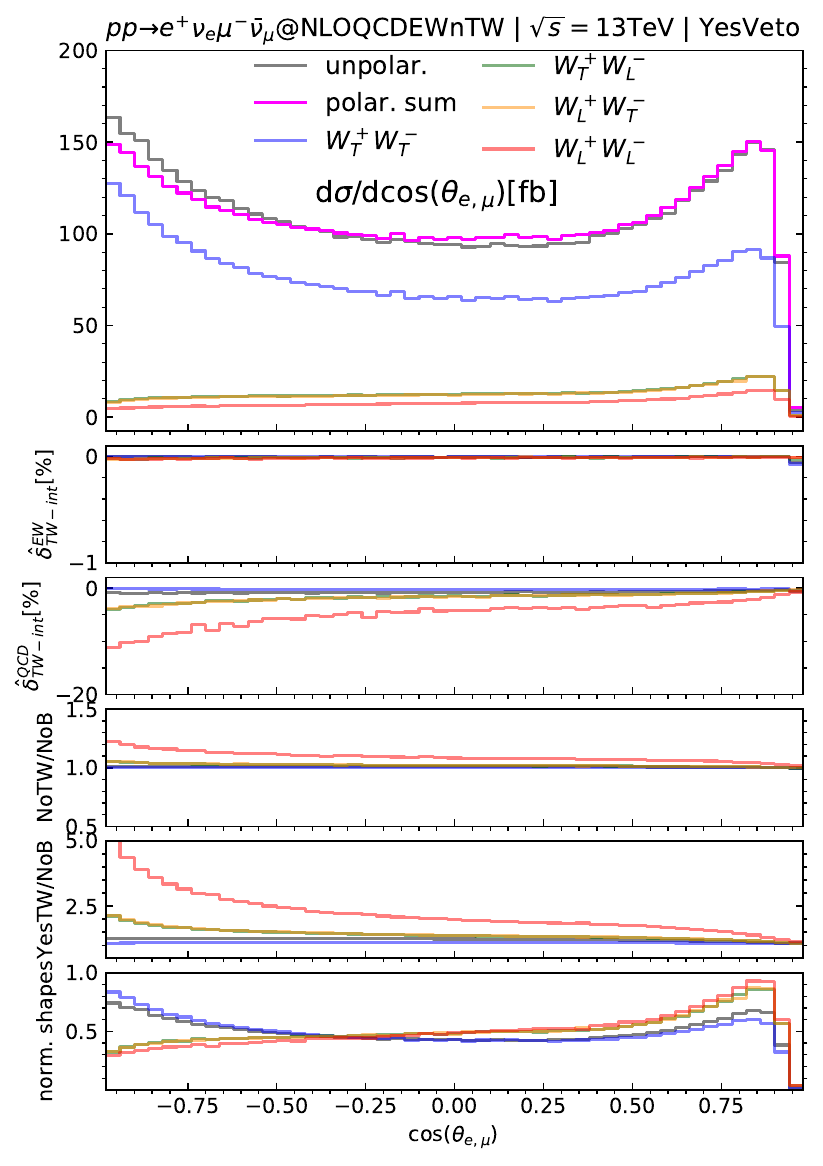} 
  \includegraphics[width=0.48\textwidth]{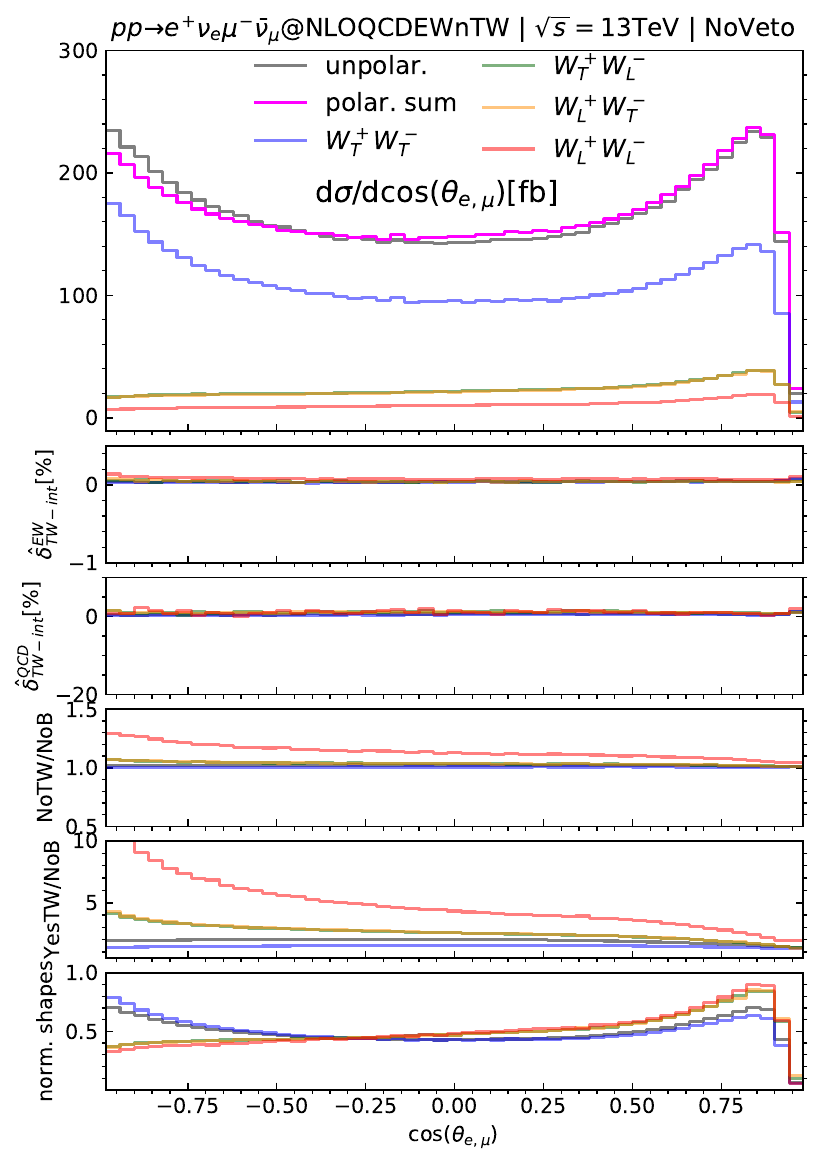}\\
  \includegraphics[width=0.48\textwidth]{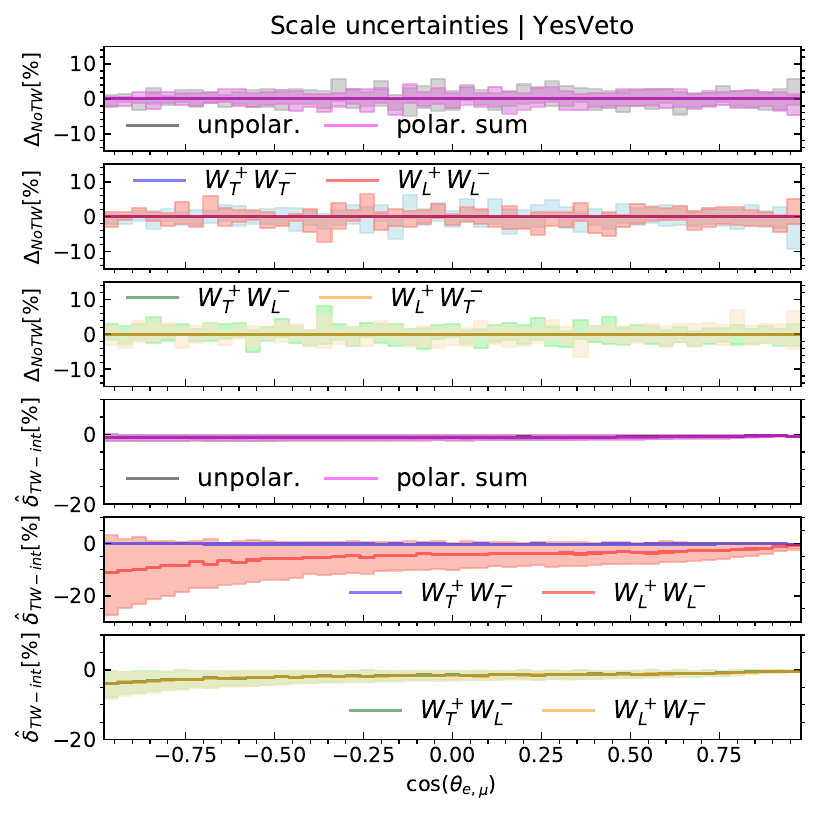}
  \includegraphics[width=0.48\textwidth]{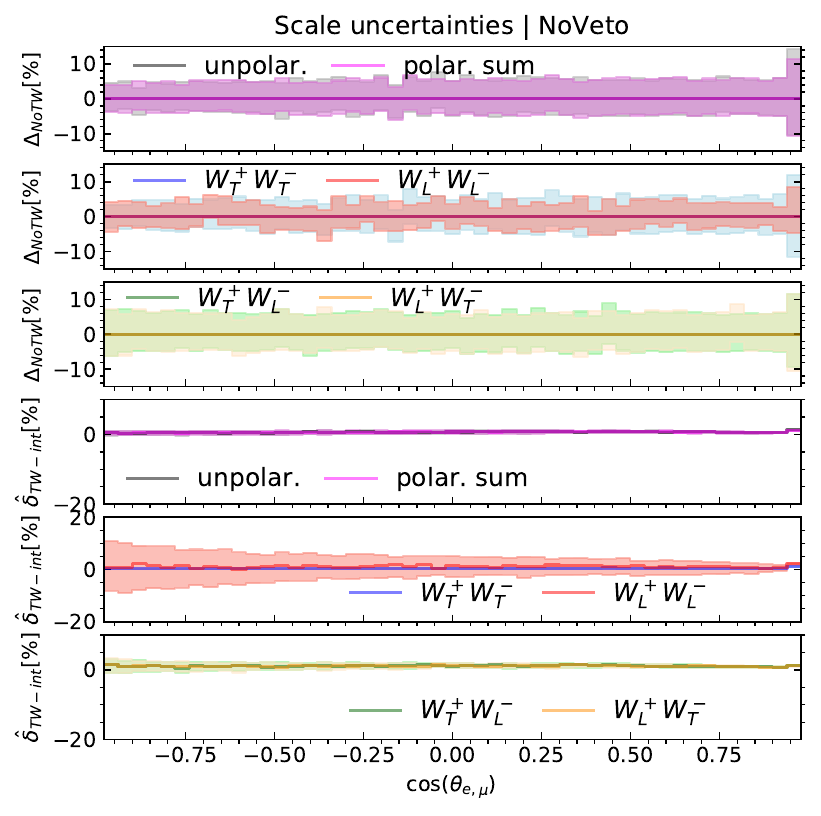}
  \end{tabular}
  \caption{Same as \fig{fig:dist_costheta_e_VV} but for the $\cos(\theta_{e^+,\mu^-})$ variable.}
  \label{fig:dist_cos35}
\end{figure}

\begin{figure}[t!]
  \centering
  \begin{tabular}{cc}
  \includegraphics[width=0.48\textwidth]{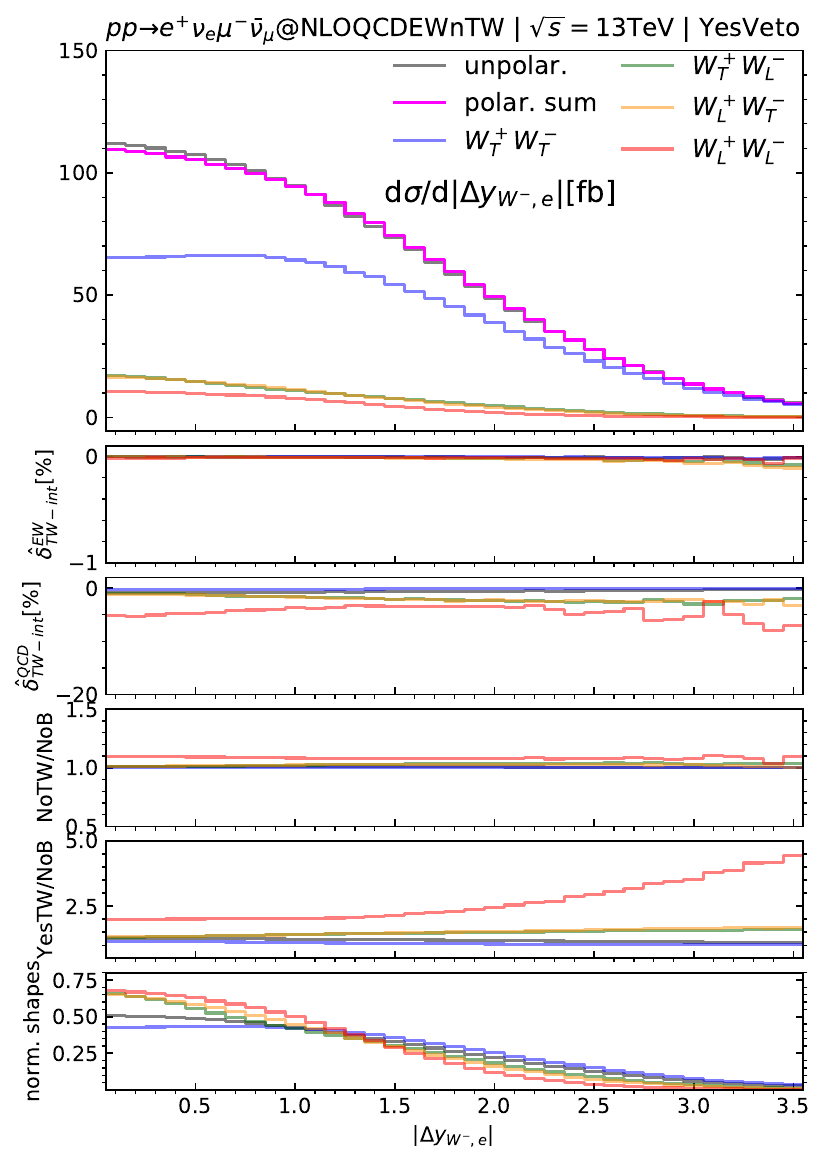} 
  \includegraphics[width=0.48\textwidth]{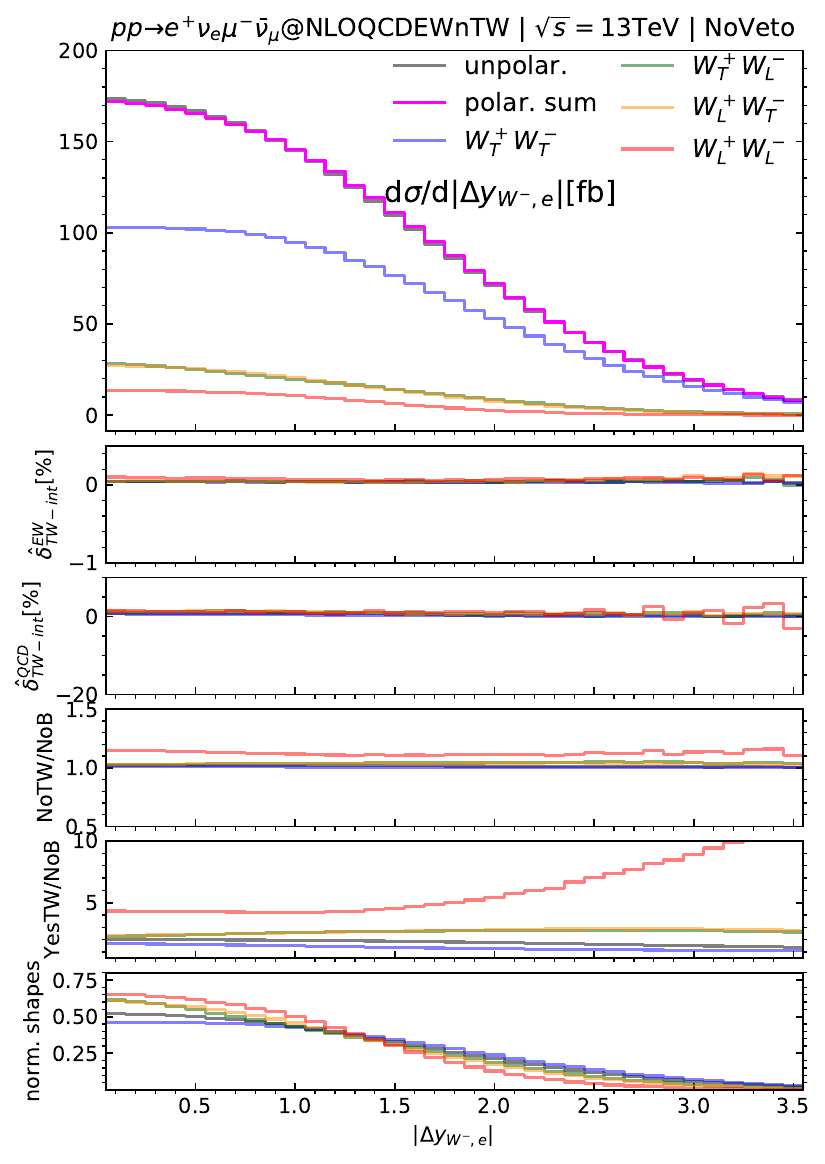}\\
  \includegraphics[width=0.48\textwidth]{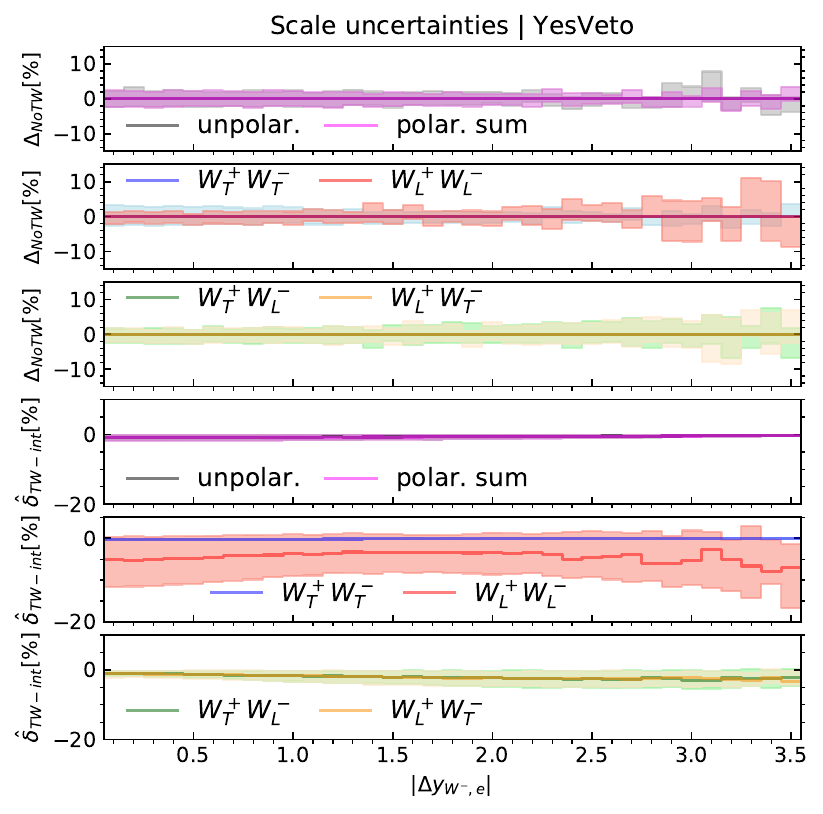}
  \includegraphics[width=0.48\textwidth]{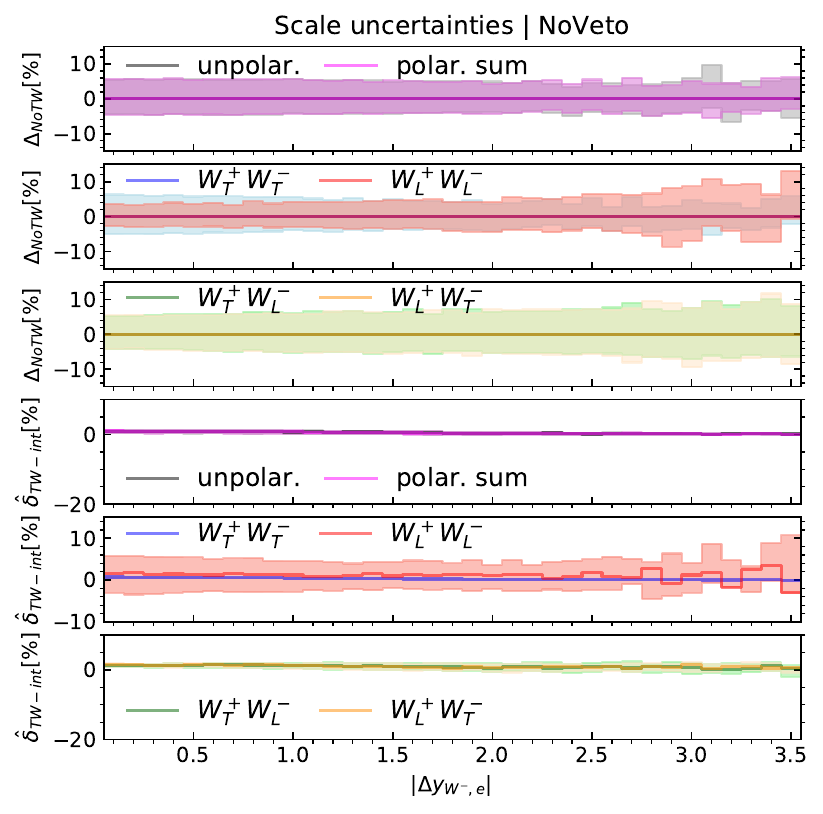}
  \end{tabular}
  \caption{Same as \fig{fig:dist_costheta_e_VV} but for the rapidity separation between the positron and the $W^-$ (absolute value).}
  \label{fig:dist_Dy}
\end{figure}

\begin{figure}[t!]
  \centering
  \begin{tabular}{cc}
  \includegraphics[width=0.48\textwidth]{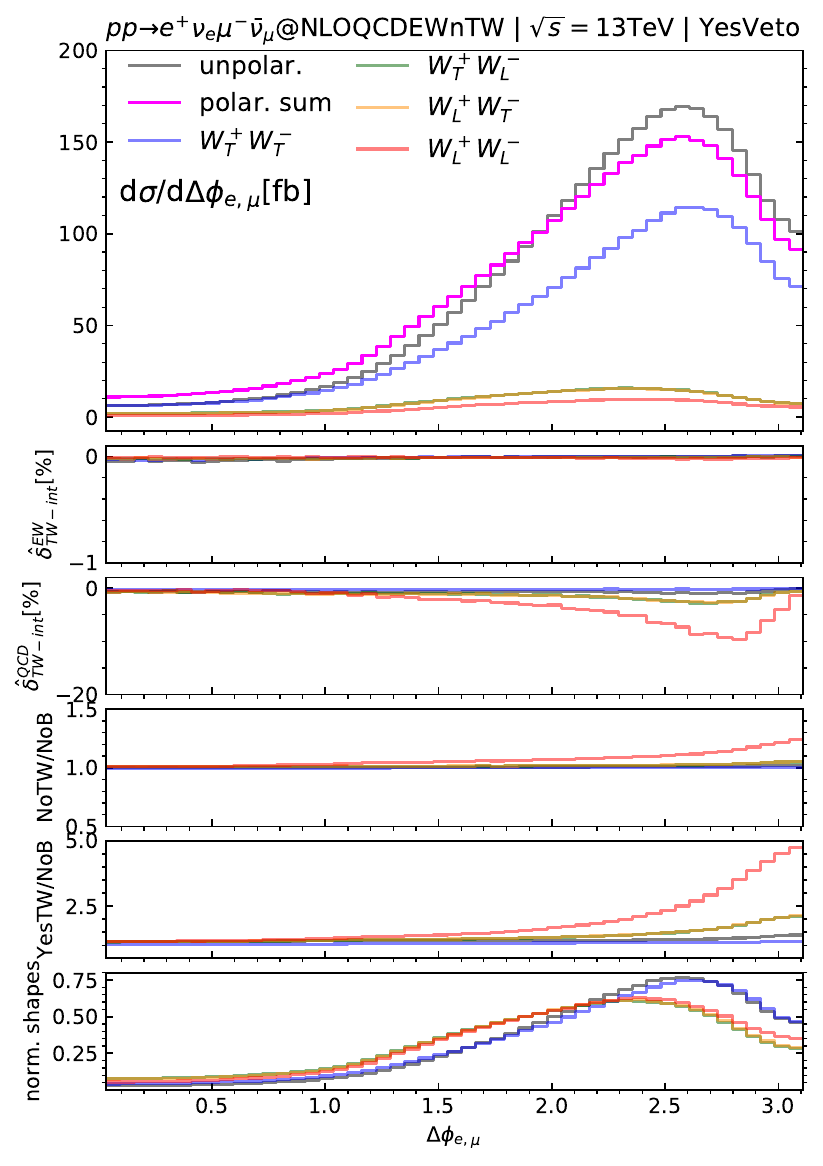} 
  \includegraphics[width=0.48\textwidth]{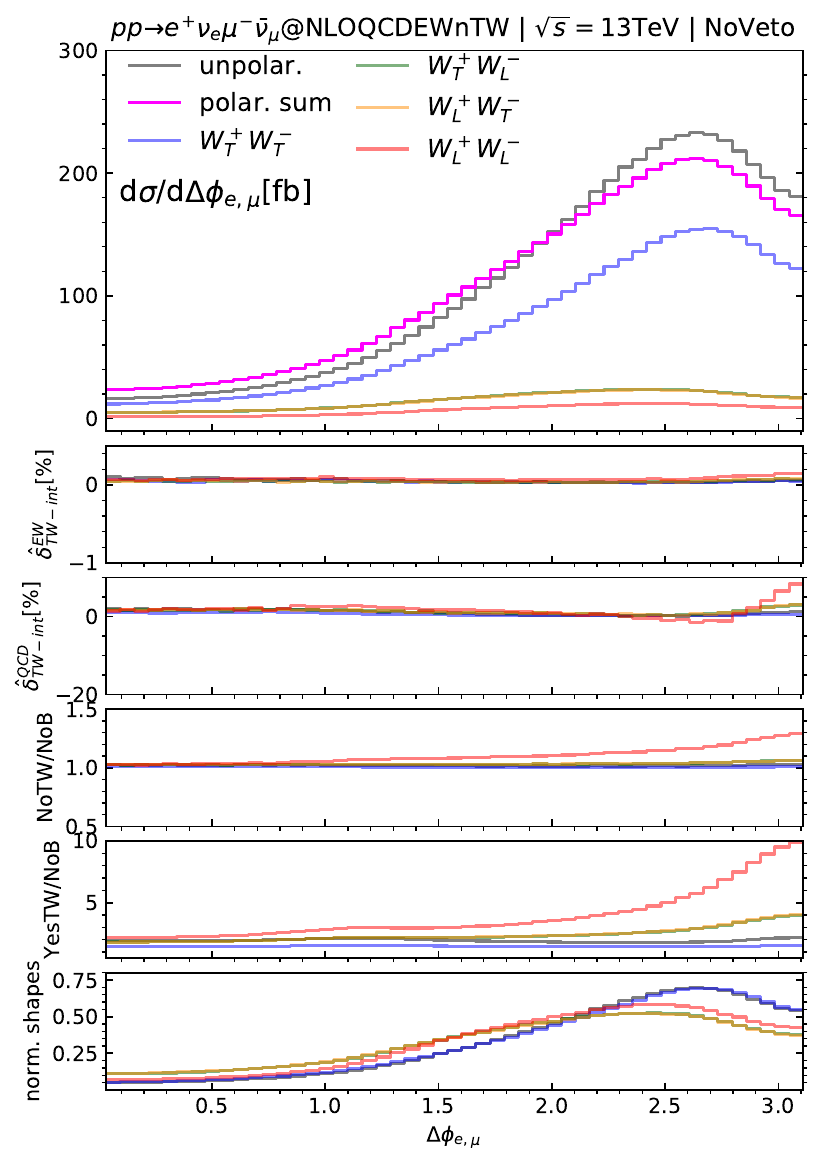}\\
  \includegraphics[width=0.48\textwidth]{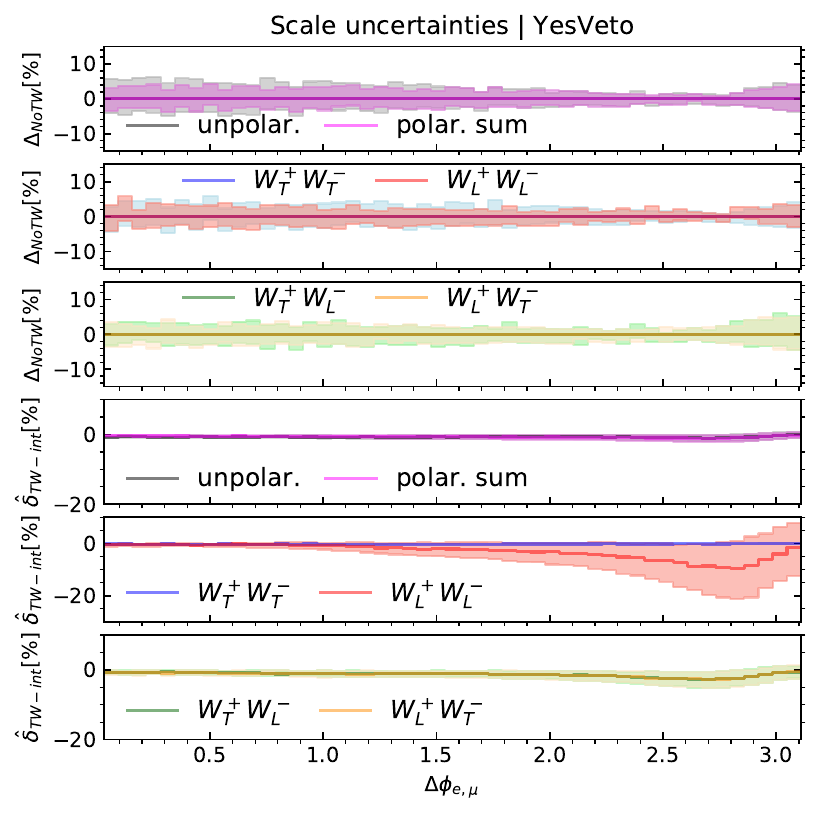}
  \includegraphics[width=0.48\textwidth]{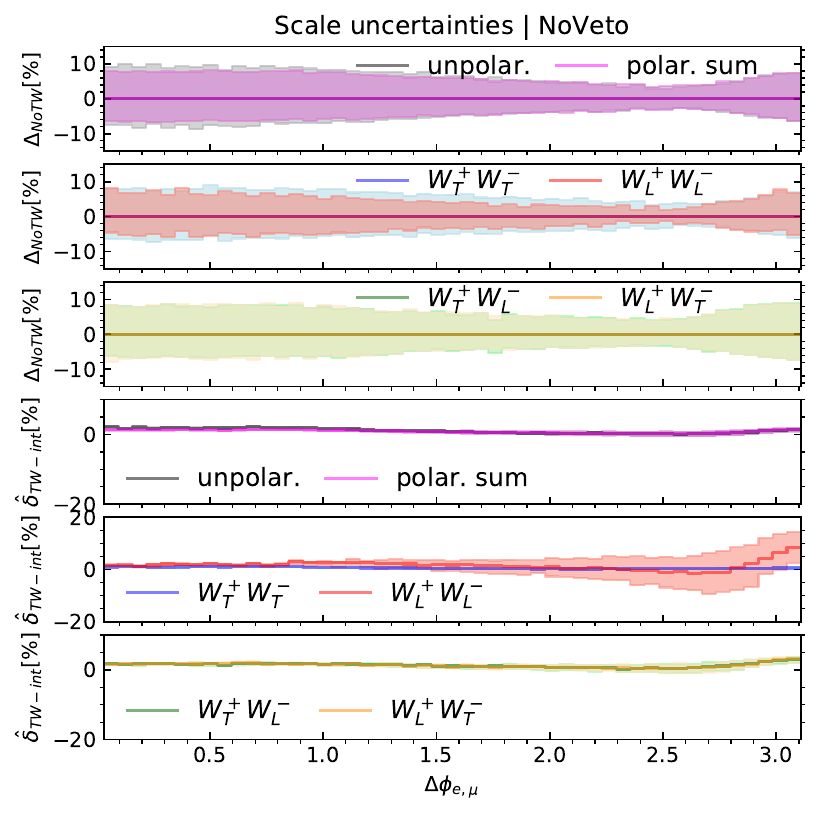}
  \end{tabular}
  \caption{Same as \fig{fig:dist_costheta_e_VV} but for the azimuthal-angle separation between the positron and the muon.}
  \label{fig:dist_Dphi}
\end{figure}

\begin{figure}[t!]
  \centering
  \begin{tabular}{cc}
  \includegraphics[width=0.48\textwidth]{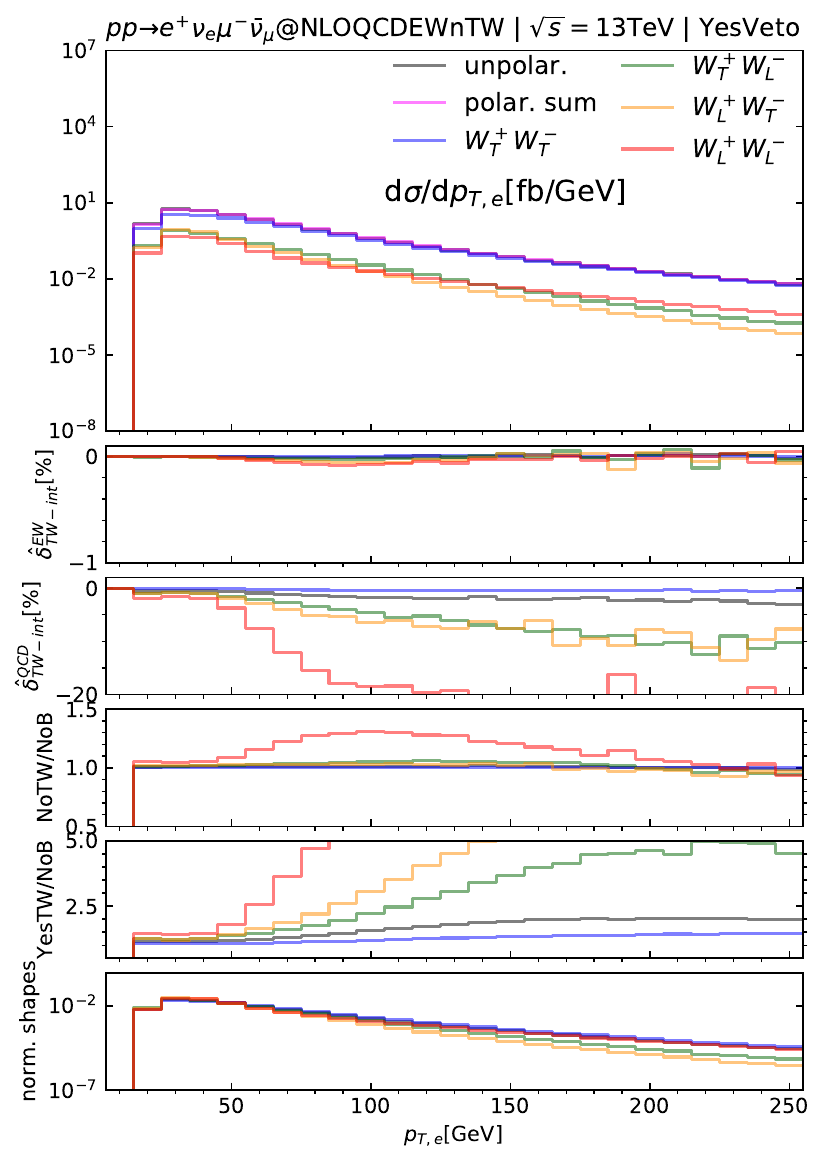} 
  \includegraphics[width=0.48\textwidth]{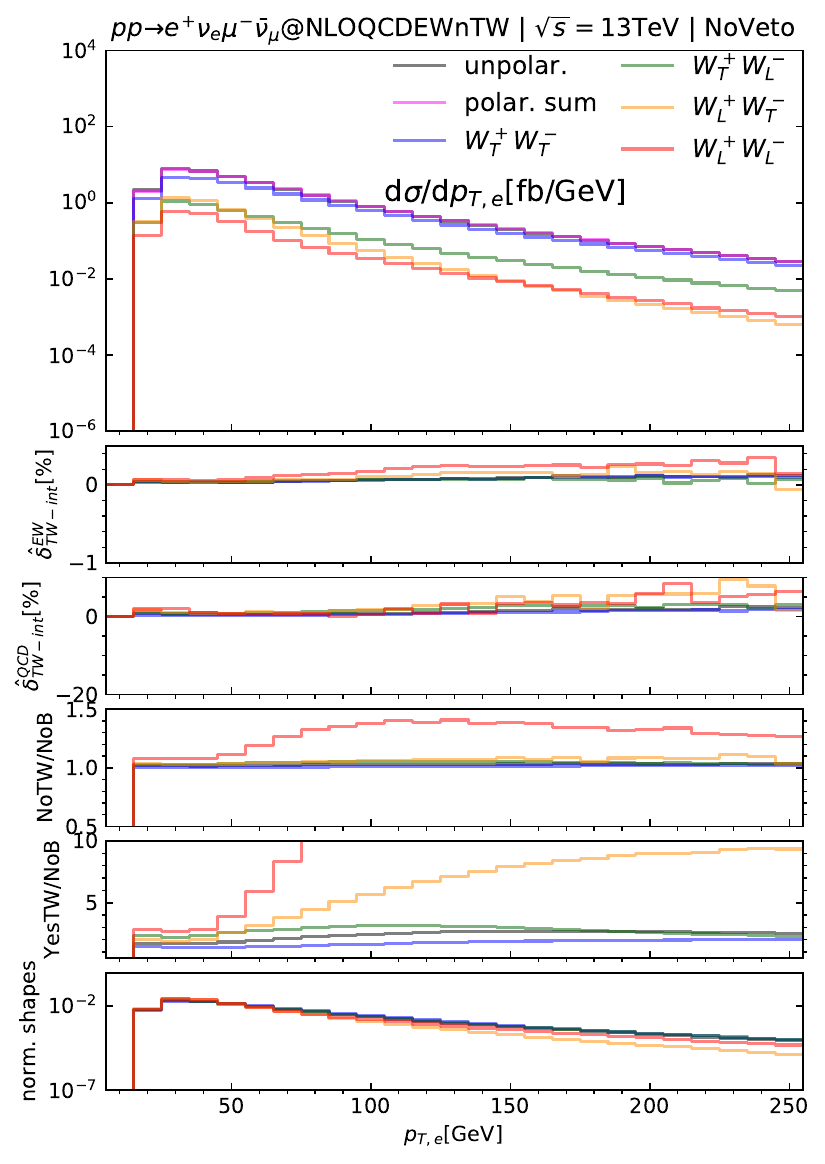}\\
  \includegraphics[width=0.48\textwidth]{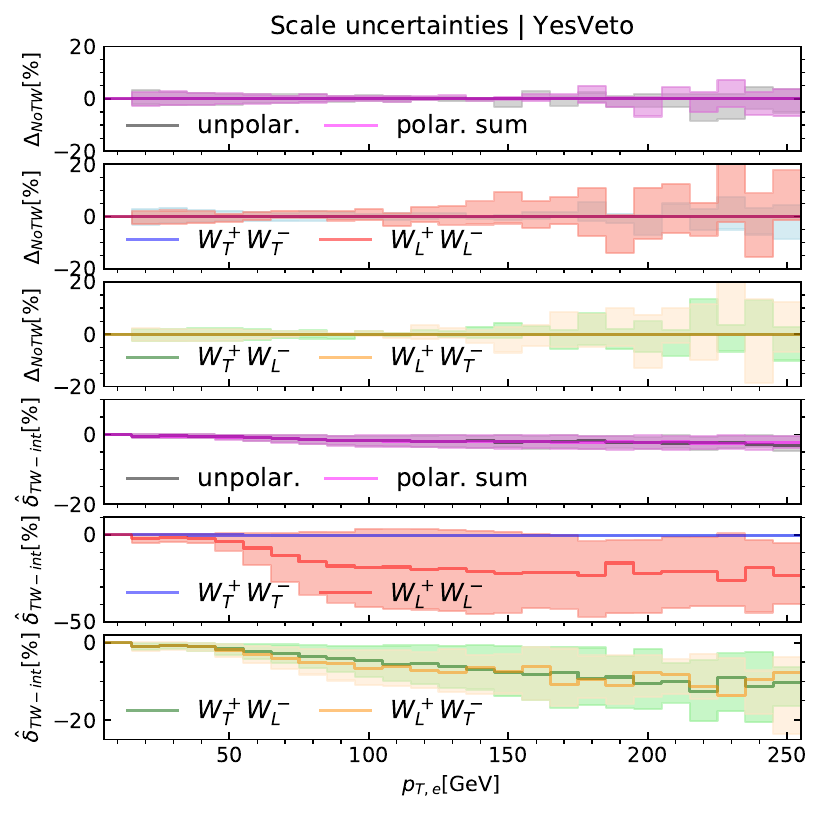}
  \includegraphics[width=0.48\textwidth]{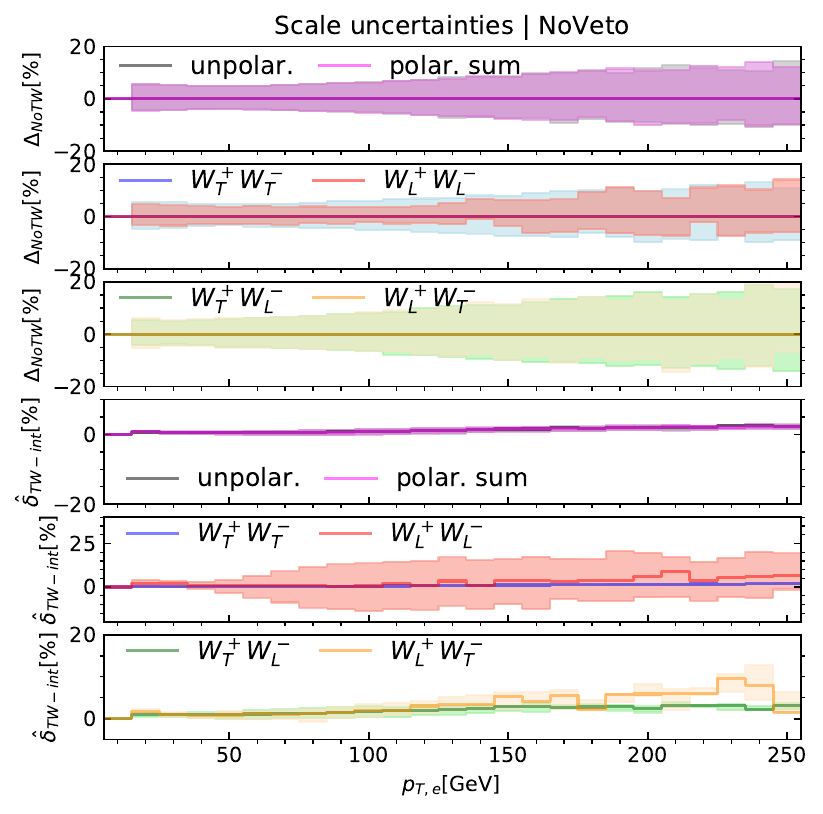}
  \end{tabular}
  \caption{Same as \fig{fig:dist_costheta_e_VV} but for the transverse momentum of the positron.}
  \label{fig:dist_pT_e}
\end{figure}

We provide here additional kinematic distributions, with the same notation, format, and color code as 
the ones presented in \sect{sect:dist}. All kinematic variables ($X$ axis) are calculated in the laboratory frame.
The distribution in $\cos(\theta_{e^+,\mu^-})$, the angle between the two leptons, 
is shown in \fig{fig:dist_cos35}. 
The distribution in the rapidity separation between the positron and the $W^-$, $|\Delta y_{W^-,e^+}| = |y_{e^+} - y_{W^-}|$, 
is presented in \fig{fig:dist_Dy}.  
The distribution in the azimuthal-angle separation between the positron and the muon is displayed in \fig{fig:dist_Dphi}. 
Finally, the positron's transverse momentum distribution is given in \fig{fig:dist_pT_e}.
\acknowledgments
We would like to thank the anonymous referee of our previous paper \cite{Dao:2023kwc} for asking a 
very good question on the irreducible top-quark backgrounds, which encouraged us to undertake 
the investigation presented in this work.  
We are grateful to Ansgar Denner and Giovanni Pelliccioli for helpful discussions and providing us results 
for the comparison with \bib{Denner:2023ehn}.
This research is funded by Phenikaa University under grant number PU2023-1-A-18.


\providecommand{\href}[2]{#2}\begingroup\raggedright\endgroup
\end{document}